\documentclass[aps,prx,showpacs,twocolumn,superscriptaddress,longbibliography]{revtex4}
\usepackage[utf8]{inputenc}
\usepackage[american,british]{babel}
\usepackage[T1]{fontenc}
\usepackage[pdftex]{graphicx}  
\usepackage{graphicx, xcolor}
\usepackage{dcolumn}
\usepackage{bm}
\usepackage{amsmath,amsthm,amssymb}
\usepackage{color}
\usepackage{verbatim}

\definecolor{darkGreen}{RGB}{0,110,0}
\definecolor{darkBlue}{RGB}{0,0,130}
\usepackage[colorlinks,citecolor=darkGreen,linkcolor=darkBlue,%
urlcolor=blue,hyperindex]{hyperref}

\usepackage{ulem} 

\newcommand{\be}{\begin{equation}}
\newcommand{\ee}{\end{equation}}

\footnotetext{Equal contribution.}

\usepackage{bm}

\begin{document}

\title{Unsupervised learning universal critical behavior via the intrinsic dimension}

\author{T. Mendes-Santos$^*$}
\affiliation{The Abdus Salam International Centre for Theoretical Physics, strada Costiera 11, 34151 Trieste, Italy}
\author{X. Turkeshi$^*$}
\affiliation{The Abdus Salam International Centre for Theoretical Physics, strada Costiera 11, 34151 Trieste, Italy}
\affiliation{SISSA,via Bonomea, 265, 34136 Trieste, Italy}
\affiliation{INFN,via Bonomea, 265, 34136 Trieste, Italy}
\author{M. Dalmonte}
\affiliation{The Abdus Salam International Centre for Theoretical Physics, strada Costiera 11, 34151 Trieste, Italy}
\affiliation{SISSA,via Bonomea, 265, 34136 Trieste, Italy}
\author{Alex Rodriguez}
\affiliation{The Abdus Salam International Centre for Theoretical Physics, strada Costiera 11, 34151 Trieste, Italy}

\begin{abstract}
The identification of 
universal properties from minimally processed data sets is one goal of machine learning techniques applied to statistical physics.
Here, we study how the minimum number of variables needed to accurately describe the important features of a data set - the intrinsic dimension ($I_d$) - behaves in the vicinity of phase transitions. We employ state-of-the-art nearest neighbors-based $I_d$-estimators to compute the $I_d$ of raw Monte Carlo thermal configurations across different phase transitions: first-, second-order and Berezinskii-Kosterlitz-Thouless.
For all the considered cases, we find that the $I_d$ uniquely characterizes the transition regime. The finite-size analysis of the $I_d$ allows not just to identify critical points with an accuracy comparable with methods that rely on {\it a priori} identification of order parameters, but also to determine the corresponding (critical) exponent $\nu$ in case of continuous transitions. For the case of topological transitions, this analysis overcomes the reported limitations affecting other unsupervised learning methods. Our work reveals how raw data sets display unique signatures of universal behavior in the absence of any dimensional reduction scheme, and suggest direct parallelism between conventional order parameters in real space, and the intrinsic dimension in the data space. 
\end{abstract}

\maketitle
\section{Introduction} 

\begin{figure*}[t]
\begin{center}
\includegraphics[width=0.95\textwidth]{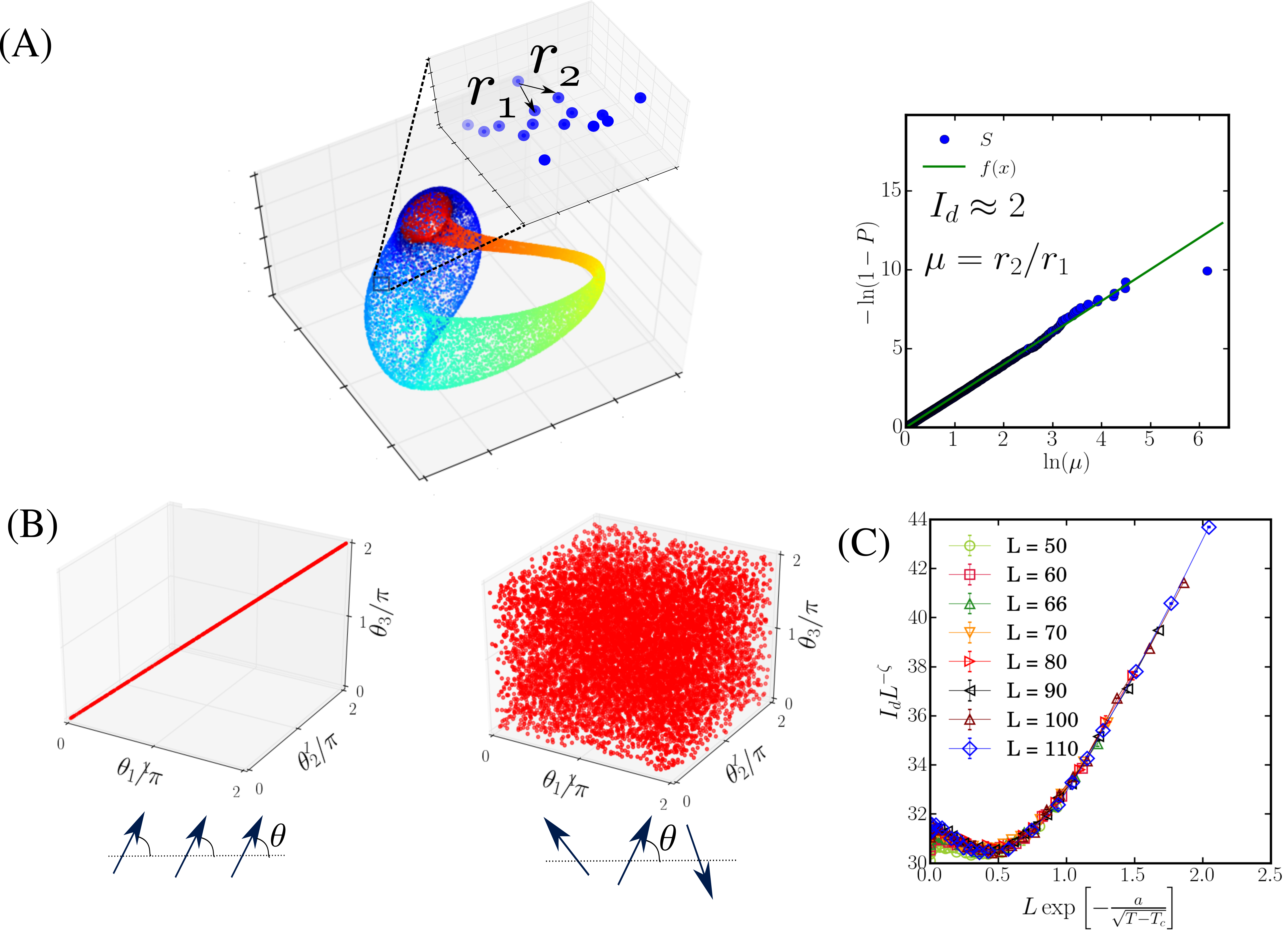}
\end{center}
\caption{Panel (A): \textit{Schematics of the intrinsic dimension, $I_d$.}
The important content of a data set typically lays in a manifold whose $I_d$ is much lower than the number of coordinates.
In the example, despite the synthetic data set (Klein's bottle-shaped) is embedded in a three-dimensional space, it can be effectively described by a twisted manifold whose ${I_d = 2}$.
The key ingredients to compute the $I_d$ are the first- and the second-nearest neighbor distances, $r_1$ and $r_2$, of each point of the data set. 
The computation of the $I_d$ is based on the fitting of the \textit{empirical} cumulative distribution function (CDF) of the ratio ${\mu = r_2/r_1}$, $P(\mu)$ [see text and Eq.~\eqref{Id}].
Panels (B): \textit{Low- and high-temperature data sets of a 3-site model in configuration space.}
The points represent the $3$-site XY model configurations: ${\vec{\theta} = (\theta_1, \theta_2, \theta_{3})}$.
The high- and zero-temperature cases show simple data structures: for ${T = 100}$, $I_d$ is equal to the number of spins, while for ${T = 0}$, ${I_d = 1}$.
Panel(C): \textit{Intrinsic dimension in the vicinity of a phase transition.}
The $I_d$ in the intermediate temperature regime, representative of phase transitions in larger systems, is considerably more complex.
The temperature dependence of $I_d$ can be used to signalize and characterize critical points. As an example, we show the universal data collapse of the $I_d$ at the Berezinskii-Kosterlitz-Thouless described 
by the 2D XY model.
}
\label{fig1} 
\end{figure*}

The growing field of machine learning (ML) is rapidly expanding our capabilities of analyzing and describing high-dimensional data sets~\cite{Jordan255,lecun2015deep,domingos2012few,butler2018machine}. With the increasing understanding of these methods, the community is becoming convinced that their outstanding performance is mostly due to the fact that this 
``high-dimensionality'' is applicable only to the embedding space, while the data sets lay in a manifold that can be twisted and topologically complex but whose intrinsic dimension, $I_d$, is typically much smaller than the large number of coordinates of the system~\cite{Bickel2005,goldt2019modelling}(see graphics in Fig. \ref{fig1} (A)).
The determination of this $I_d$ is an active field of research~\cite{Rozza2015,Bickel2005,Laio2017} in unsupervised learning (UL), i.e., the branch of machine learning that aims to uncover the internal structure of a data set without the need of any label.

Recently, ML ideas have encountered fruitful applications in the context of statistical physics~\cite{Carleo2019,Pankaj2019,Carrasquilla2020}. Such applications have ranged from the determination of physical properties~\cite{LeiWang2016,Melko2017,Huber2017,Kim2017,Eshan2017,Benno2019,Annabelle2019,Zhang2019,Lucini2020}, to the formulation of novel classes of variational ans\"atze~\cite{Carleo2017,DasSarma2017,Melko2018bm,Markus2019}. These progress leveraged on analyzing and exploiting the results of dimensional reduction, and using a variety of tools to analyze (or employ) the final representation (or truncation) obtained in this way. In various contexts, results obtained via these methods have remarkably shown to be already competitive with more traditional approaches~\cite{Carleo2019}. 

Here, we pursue an alternative approach: our main purpose is to show that, from a ML perspective, physically relevant and universal information can be gathered by analyzing the very same embedding procedure that carries out the dimensional reduction, rather than focusing on its final result. In particular, we show how the intrinsic dimension correspondent to the partition function of statistical mechanics models displays universal scaling behavior in the vicinity of phase transitions, and it behaves as an order parameter for a corresponding structural transition in data space. Differently from previous works~\cite{LeiWang2016,Wetzel2017,Eshan2018,Scalettar2017,Singh2017,Scheurer2019,Yang2020,Chinesta2018}, our approach is thus focused on {\it data mining} the data set as a whole, and thus, does not leverage on any kind of projection. At the technical level, this is achieved by employing a cutting-edge nearest-neighbor estimator of the $I_d$, which is suitably designed to deal with non-linear data sets, i.e. data sets lying on non-linear manifolds~\cite{Laio2017}.

In order to access the complex data structure at phase boundaries, we study numerically instances of first-order, second-order (conformal), and Berezinskii-Kosterlitz-Thouless (BKT) 
transitions in two-dimensional (2D) classical spin systems.
In all cases, $I_d$ displays a universal scaling behavior correspondent with the transition properties of the underlying lattice model. 
(i) For first-order transitions, $I_d$ peaks at the critical point due to the coexistence of different orders, and the finite-size corrections of the transition temperature are dictated by trivial scaling exponents. 
(ii) For both second-order and topological transitions, we observe {\it universal scaling collapse}, with transition temperature and critical exponents determined to the percent level.
(iii) Most importantly, we provide compelling evidence that the $I_d$ is an ideal tool to underpin topological transitions in an unsupervised fashion: as an example, we extract the critical temperature of the 2D XY model with 1\% confidence even at modest system sizes.

We then develop a theoretical framework in support of the fact that $I_d$ has characteristic features at transition points, that are governed by scaling theory. First, we show how several instances of the data set, in particular, the distribution of distances between sampled configurations, are already revealing striking features about critical behavior for all classes of phase transitions.
The basic idea there is that the data space naturally clusters configurations characterized by similar physical properties (e.g., magnetization and winding number). 
The fact that the intrinsic dimension has strong features at phase transition then follows from its 'local' nature (related to changes of scale in configuration space, in a sense that we specify below). Then, we discuss how, for the type of data sets we are interested in, the intrinsic dimension can be connected directly to a set of arbitrary-many-body correlation functions, that, following the finite-size scaling hypothesis, justifies its scaling behavior in the vicinity of transition points. We then check {\it a posteriori} the validity of some of the assumptions at the basis of this framework.

Before diving into the main part of our manuscript, we provide a simplified picture that qualitatively captures how the intrinsic dimension is connected to the physical information obtained by sampling a partition function via Monte Carlo methods. The basic intuition behind the $I_d$ of data sets generated in the low- and high-temperature regimes of a simple $3$-site XY model is drawn in  Fig.~\ref{fig1} (B).
At low temperature (left graphics of Fig.~\ref{fig1} (B)), most of the spin configurations sampled during the Markov chain correspond to fully ferromagnetic spin arrangements (see cartoon). In the limiting case ${T = 0}$, the ground states are given by XY ferromagnetic configurations, i.e., ${\theta_1=\theta_2= \theta_{3}}$, and the data set is described by a manifold that lays in a line (${I_d = 1}$).
Oppositely, in the high temperatures regime, the data set is described by a manifold whose ${I_d = 3}$: each new Monte Carlo configuration corresponds to an arbitrary arrangement of the three spins, so that the structure of the data set is that of a homogeneously occupied three-dimensional space. This simple example demonstrate how transitions in parameter space are accompanied by structural transitions in data space. Due to its collective origin, the transition region requires the computation of $I_d$ in very high dimensional data space: in Fig. \ref{fig1} (C), we show a sample of our results, illustrating the scaling collapse of the intrinsic dimension correspondent of the 2D XY model in the vicinity of its BKT transition point.

\section{Intrinsic Dimension}  

Before addressing the analysis of concrete statistical mechanics models, we present here a self-contained discussion on the intrinsic dimension and its state-of-the-art estimators. This section is propaedeutic to the critical identification of the best estimator to be used in our applications below. 

The $I_d$ is a concept that arises from the observation that, in natural data sets, the correlations between the input variables induce a structure, modifying the dimensionality of the manifold in which the data lie. 
In order to visualize this, one can imagine a data set with the Cartesian coordinates of points extracted from a circle.
Although the input coordinates are two, they are strongly correlated and the manifold in which the points lie has a $I_d=1$. 
Therefore, in simple cases like this, or the one shown in Fig. \ref{fig1}B, it roughly corresponds with the minimum number of variables needed to describe a data set~\cite{Laio2017, Rozza2015}.

Information about the $I_d$ is important to determine if dimensional reduction of high-dimensional data sets incurs information loss or not.
Moreover, it can be used as an UL approach to characterize a system. Just to mention a few examples: in biological physics, the $I_d$ can be used to determine the number of independent directions a protein can have during a sequence evolution~\cite{Facco2019},
in image analysis, to distinguish between different kind of image structures~\cite{Krueger2003},
in astrophysics, to estimate the amount of information available in spectropolarimetric data~\cite{ramos2007intrinsic},
in theoretical machine learning, to understand the properties of deep neural networks~\cite{ansuini2019intrinsic},
and in ecology, to characterize the minimum number of independent axes of variation that adequately describes the functional variation among plants~\cite{Laughlin2014traits}.

Different approaches have been developed to estimate the $I_d$, see Ref.~\cite{Rozza2015} for review.
For example, dimensional reduction techniques, such as principal component analysis (PCA)~\cite{Geladi2001}, Multidimensional Scaling~\cite{MDS}, Isomap~\cite{ISOMAP},
Locally Linear Embedding~\cite{LLE}, Autoencoders~\cite{AE}, t-distributed stochastic neighbor embedding (t-SNE)~\cite{TSNE} or Uniform Manifold Approximation and Projection (UMAP)~\cite{lel2018umap} to mention some of them, search for a lower dimensional space to project the data set by minimizing a projection error.
The dimension of the identified subspace is viewed as an estimation of $I_d$. However, identifying this dimension is far from trivial. 
For instance, in the PCA case, one should take into consideration the spectrum of the eigenvalues of the covariance matrix and look either 
for a gap or decide {\it ad hoc} a cut-off parameter. 
It is worth saying that, for PCA, this strategy will not work if the manifold of lower dimensionality is curved. 
Furthermore some of the above mentioned methods, like t-SNE, are focused on visualization and assume that the dimension of the projection space is lower than the $I_d$. 
Therefore, these projection algorithms aim to alleviate the problems that this dimension mismatch cause in the visualization, not being well suited for $I_d$ detection.

A closely related quantity is the fractal dimension~\cite{camastra2002estimating}, whose estimation relies on the scaling of the number of neighbors with the distance from a given point. 
This approach is largely employed in the study of percolation transitions~\cite{Stauffer1991}, but it suffers from serious limitations when the density distribution of points is not uniform. 

These limitations lead to the development of nearest neighbors methods, in which it is assumed that nearest-neighborhood points can be considered as uniformly drawn 
from small enough $I_d$-dimensional hyperspheres (not all the data set)~\cite{Rozza2015,Bickel2005}.
Indeed, the avoidance of any projection step and the smoothing on the condition of data uniformity (from the full data set to 
a small neighborhood around each point) are key features for obtaining good results in highly non-uniform, non-linear data sets even at
really high dimensions (a regime at which all the purely geometrical methods present a bias due to the curse of dimensionality).

The TWO-NN method employed in this work belongs to this type of methods, with the particularity that by focusing only on the first two nearest neighbors (see Fig. \ref{fig1} A), the size of the $I_d$-dimensional hyperspheres at which the density is assumed constant is reduced to its minimum expression. The method is rooted in computing the distribution functions of neighborhood distances, which are function of $I_d$.
More specifically, for each point $\vec{x}$ in the data set, we consider its first and second nearest-neighbors distances $r_1(\vec{x})$ and  $r_2(\vec{x})$, respectively.
Under the condition that the data set is locally uniform in the range of second nearest-neighbors, it has been shown in Ref.~\cite{Laio2017} that the the distribution function of ${\mu = r_2(\vec{x})/r_1(\vec{x})}$ 
is
\begin{align}
f(\mu) = I_d \mu^{-I_d - 1}.    
\label{eq:fmu}
\end{align}
Or, in terms of the cumulative distribution, $P(\mu)$,
\begin{align}
 I_d = - \frac{\ln\left[ 1 -  P(\mu) \right]}{\ln\left( \mu \right)},
 \label{Id}
\end{align}
which can be used to obtain  $I_d$ by fitting $S = \{ (\ln(\mu), - \ln\left[ 1 - P^\textup{emp}(\mu) \right]   \}$ with a 
straight line passing through the origin.
The function  $P^\textup{emp}$ defines the empirical cumulate and is computed  by sorting the values of $\mu$ in a ascending order, see Appendix \ref{distances} for more details.
In Fig. \ref{fig1} A, the steps for computing the $I_d$ in a highly non-linear manifold with complex topology (in this case, a Klein's bottle-shaped data set) are summarized: a) Compute the distance from the first and second neighbors b) Compute for each point $\mu$ and its empirical cumulate and c) fit $S$ to a straight line. 

We stress that this method is not free of drawbacks. As mentioned above, being a purely geometrical method, it is affected by the curse of dimensionality, since the number of points needed to have an accurate measure of the $I_d$ grows exponentially with the $I_d$. 
Moreover, Equation (\ref{Id}) was derived assuming a continuous real support. Therefore, applying it to data sets with a different support implies some degree of approximation that can fail in some limiting cases. For instance, this shall happen when two or more configurations have the same coordinates.  However, as we detail below, these drawbacks do not affect the results obtained in this work: in particular, these limitations do not kick in when investigating transitions, even when configuration spaces are composed of discrete variables such as Ising spins. These limitations only affect data sets corresponding to either very small system sizes, or phases at extremely low temperatures where, during the MC sampling, configurations may be repeated as the accessible configuration space is very limited.

\section{Models}

Our approach focuses on the high-dimensional data sets associated with the equilibrium configuration states of a partition function.
Such states are sampled with Markov Chain Monte Carlo simulations from the thermal weight $\rho(E) \sim e^{-E(\vec{x})/T}$,
where $E(\vec{x})$ is the energy of an independent configuration $\vec{x}$ and $T$ is the temperature.
We employ Wolff's cluster algorithm~\cite{Wolff1989,Landau2005}, and for each data set, we consider $N_r$ configurations.

We consider partition-function data sets of several models in the vicinity of various types of phase transitions~\cite{DiFrancesco1997,Henkel1999}.
The first example is the well known Ising model in two-dimensions
\begin{align}
 E(\vec{s}) = - \sum_{\left< i,j \right>} s_i s_j,
 \label{Ising}
\end{align}
where the spin degrees of freedom are $s_i = \pm 1$, and  $\left< i,j \right>$ are the nearest neighboring bonds of a square lattice, with  $N_s = L \times L$ spins and periodic boundary condition. 
The Ising configuration states are defined as
\begin{equation}
\vec{s} = (s_1, s_2, ..., s_{N_s}).
\label{confIsing}
\end{equation}
This model describes a second-order  phase transition characterized by the breaking of a  $Z_2$ symmetry at the critical temperature $T_c = 2/\ln(1 + \sqrt{2})$. 
In the vicinity of $T_c$, the spin correlation length diverge as $\xi \sim (T - T_c)^{-\nu}$, where the critical exponent is $\nu = 1$.

We also consider the first- and second-order phase transitions described by the $q$-states Potts model (qPM)  
\begin{align}
 E(\vec{\sigma}) = - \sum_{\left< i,j \right>} \delta_{\sigma_i,\sigma_j},
 \label{qPM}
\end{align}
where the spin $\sigma_i = 0,1,2, ..., q - 1$,
and $\delta_{\sigma_i,\sigma_j}$ is the delta function.
In particular, the ${q = 2}$ Potts model can be mapped into the Ising model.
The Potts  configuration states are defined by
\begin{equation}
\vec{\sigma} = (\sigma_1, \sigma_2, ..., \sigma_{N_s}).
\label{confPotts}
\end{equation}
The qPM is characterized by a discrete $Z_q$ symmetry that is broken at the critical temperature $T_c = 1/\ln(1 + \sqrt{q})$. Importantly, this class of models displays a second-order phase transition for $q\le 4$, and a first-order one for $q>4$. We examine both these regime: the second-order transition described by the ${q=3}$ PM (with correlation length critical exponent $\nu=4/5$), and the first-order transition described by the ${q = 8}$ PM~\cite{Wu1982,sandvik2019}.

Finally, as a representative of the BKT universality class, we investigate the two-dimensional XY model\cite{Kosterlitz1973,Batrouni1988,nori2020}
\begin{align}
 E(\vec{\theta} ) = - \sum_{\left< i,j \right>} \vec{S}_{i} \cdot \vec{S}_{j},
 \label{xymodel}
\end{align}
where $\vec{S}_{i} = (\cos(\theta_i),\sin(\theta_i))$, $\cos(\theta_i)$ and  $\sin(\theta_i)$ being the projection of the spin at site $i$ in the $x$ and $y$ directions, respectively, and  $\theta_i$ $\in$ $\left[0,2\pi \right[$.
The XY configurations are defined as
\begin{equation}
 \vec{\theta} =[\cos(\theta_1), \sin(\theta_1),...,\cos(\theta_{N_s}), \sin(\theta_{N_s})].
 \label{confXY}
\end{equation}
This model is characterized by a continuous $U(1)$ symmetry and 
describes a phase transition between a high-temperature phase with exponentially decaying spin correlations, and a low-temperature quasi-ordered phase characterized by power-law decaying correlations.
The BKT critical temperature, $T_{BKT}$, is not known exactly;
state-of-the-art estimations based on the analysis of the spin stiffness of lattices of order $O(10^6)$ spins give  $T_{BKT} = 0.8935(1)$~\cite{sandvik2013}.

The detection of the BKT critical point is hindered by the fact that
it cannot be characterized by conventional local order parameters, as in the examples discussed previously, and due to the exponential growth of the correlation length near $T_{BKT}$.
Hence, the BKT transition represents a key challenging test for any UL method.

\subsection{How to characterize partition functions as data sets}

Before proceeding to the discussion of the results, we  point out some important aspects of the Ising, Potts and XY data sets (see Eqs. \eqref{confIsing}, \eqref{confPotts} and \eqref{confXY}, respectively.).
First, a crucial step to obtain the $I_d$ (cfr. Eq.~\eqref{Id}) is to consider a proper metric;
the distance $r(\vec{x^i},\vec{x^j})$ between two configuration states $\vec{x^i}$ and $\vec{x^j}$ must
be non-negative, equal to zero only for identical configurations, symmetric, and satisfy the triangular inequality.

For the XY data sets the distance is defined as the the Euclidean distance: 
\begin{equation}
r(\vec{\theta^i},\vec{\theta^j}) = \sqrt{2 \sum_{k=1}^{N_s} \left(1 - \vec{S}_{k}^{i} \cdot \vec{S}_{k}^{j} \right)}.    
\label{diseuclidian}
\end{equation}
This distance properly takes into account the periodicity of the configuration states in the interval $\theta_i$ $\in$ $\left[0,2\pi \right[$.

For both Ising and Potts  configuration states, we consider the Hamming distance, i.e.,
$r(\vec{s^i},\vec{s^j})$ (or $r(\vec{\sigma^i},\vec{\sigma^j})$) is given by the number of positions in the state vectors ($\vec{s^i}$ and $\vec{s^j}$) for which the corresponding coordinates  are different. 
The choice of the Hamming distance is motivated by the fact that the energy difference between two spins in the model of interest is given by a delta function.

As mentioned in the previous section, the two-NN method 
fails when two or more sampled configurations of the data set have identical coordinates.
This issue typically occurs in the discrete-variables Ising and Potts data sets, when
the total number of independent configuration states, $N_c$, is smaller or of the same order of the number of configurations used in the data set, $N_r$.
For instance, for both Ising and Potts data sets,  identical ferromagnetic configurations are sampled in most of the Monte Carlo steps when $T \ll T_c$.
However, in the regime that we focus here (i.e., $T$ close to $T_c$ and $L > 10$), as $N_c \gg N_r$, this issue is irrelevant (we have explicitly checked this in our data sets).

Finally, we mention that data sets generated by the XY configuration states lie typically in nonlinear manifolds,
which  can be noted by the fact that linear dimension reduction methods, such as PCA, fails to describe XY data sets, see Ref. \cite{Scheurer2019}.
In fact, even for the simple data set shown in Fig.\ref{fig1} (B), linear PCA fails in estimating the true $I_d$ of the system when the proper distance between the configurations are
taken into account, see the Appendix \ref{distances}.
This feature of the XY data sets reveals the necessity of using state-of-the-art $I_d$-estimators (such as the two-NN method considered here) that properly takes into account nonlinearities.

\section{Results}
\label{sec:results}

\begin{figure*}[t]
\begin{center}
{\centering\resizebox*{17cm}{!}{\includegraphics*{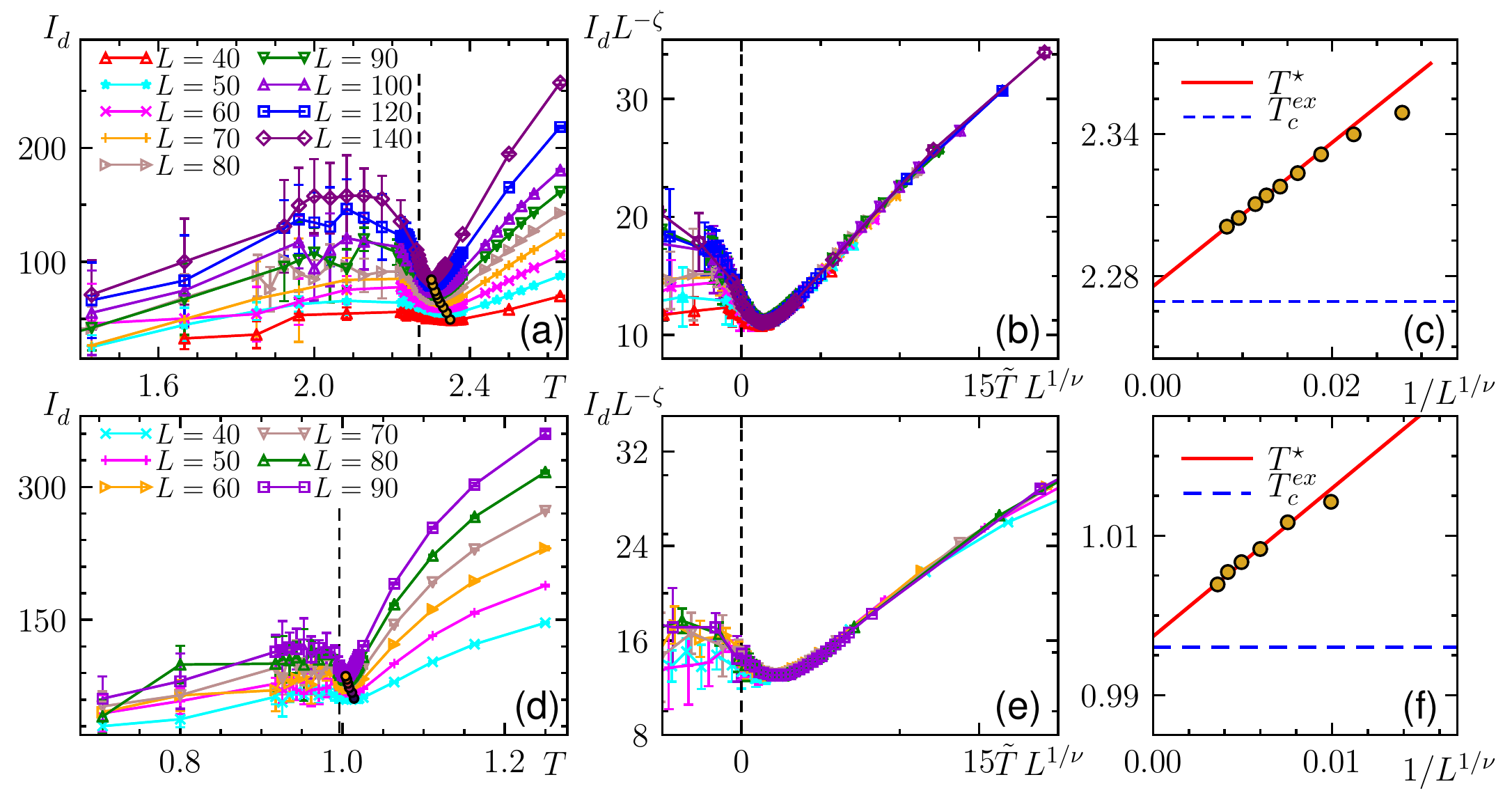}}}
\end{center}
\caption{\textit{Second-order phase transition. Panels (a-c): Ising model. Panels (d-f): $q=3$ Potts model.} Panels (a,d):  $I_d$ as a function of $T$ for the Ising model. Errorbars are standard deviation associated to a distributions of $n$ realizations of $I_d$ (we typically consider $n \ge 5$). Panel (b,e): data collapse of $I_d$ based on the FSS discussed in main text. The best data collapse of the results  give  $T_{c} = 2.283(2)$, $\nu = 1.02(2)$  and $\zeta = 0.410(5)$ for the case of the Ising model, and $T_c = 0.996(2)$, $\nu = 0.805(5)$, and $ \zeta = 0.420(2)$ for the case of the Potts model.
Panels (c,f): finite size (FSS) scaling of the minimum temperature $T^*$ (see text); the horizontal line is the exact result for $T_c$. The extrapolation returns $T_c=2.2784(2)$ and $T_c=0.9970(3)$ for the Ising and Potts cases, respectively.
}
\label{fig2}
\end{figure*}

\subsection{Second-order phase transitions}

We start our discussion considering second-order phase transitions (2PTs) described by the Ising  and the $3$-states Potts (3PM) models, see Fig. \ref{fig2}.
We consider data sets formed by $N_r = 5 \times 10^{4}$ configuration states.
Overall, far from the transition, $I_d$ is an increasing function of $T$.
For low-$T$, the computation of $I_d$ is affected by the discreteness of the Potts (Ising) configurations;
which reflects on the larger error bars. However, this issue is mitigated close to the critical point, $T_c$.
Remarkably, $I_d$ exhibits a  non-monotonic behavior in the vicinity of the critical point (see Figs. \ref{fig2} (a1) and (b1))
which can be used to locate and characterize the transition point itself.

As conventionally done in the analyses of physical observables, e.g., magnetic susceptibility and heat capacity,
we now consider a finite-size scaling (FSS) theory for $I_d$.
First, based on the FSS hypothesis and postulating that $I_d$ behaves as an order parameter for the transition, one has  $I_d = L^{\zeta} f(\xi/L)$,
where the correlation length diverges as $\xi \sim (T - T_c)^{-\nu}$, $\nu$ is a critical exponent,
and $\zeta$ is a scaling exponent associated with  the divergence of $I_d$ at $T_c$.
Figs. \ref{fig2} (a2) and (b2) show the universal data collapse for Ising and 3PM, respectively.
The values obtained for $T_c$ and $\nu$ 
have a discrepancy with exact results of less than $0.5 \%$ and $4 \%$, respectively.
For Ising we obtain: $T_c = 2.283(2)$, $\nu = 1.02(2)$, and $ \zeta = 0.410(5)$ , while for 3PM: $T_c = 0.996(2)$, $\nu = 0.805(5)$, and $ \zeta = 0.420(2)$.
See the Appendix \ref{collapse}, for the discussion about the details of the data-collapse procedure.

Further, we consider the size scaling of the shift of the local minimum of $I_d(T)$  (i.e., the temperature $T^*(L)$),
\begin{equation}
 T^*(L) - T_c \sim \frac{1}{L^{1/\nu}}.
 \label{FSS2}
\end{equation}
We note that $T^*(L)$ is reminiscent of the universal scaling behavior of singular features of physical observables close to $T_c$ (e.g., the peak of the magnetic susceptibility) \cite{sandvik2010}.
In order to compute $T_c$, we employ the following procedure: 
(i) we obtain $T^*(L)$ by fitting the results in an interval close to $T^*(L)$ with a cubic function;
the fitting is performed with a jackknife procedure, which allows establishing an error bar for $T^*(L)$.
We then (ii) consider the aforementioned FSS to compute $T_c$; 
the fitting is performed considering different sets of points.
This method provides  a coherent error propagation for $T_c$.
We obtain,  $T_c =  2.2784(2)$ for the Ising model and $0.9970(3)$ for the 3 state Potts model. Their discrepancies with the exact values are respectively of order $0.4 \%$ and $0.2 \%$.

The results of this analysis confirm the validity of our original assumption, that is, that the intrinsic dimension is a valid order parameter describing the transition in data space as a structural transition. We remark that this is validated at two steps - firstly, via the quality of the scaling collapse, and secondly, by the scaling of the transition temperature obtained by analysis a single feature of the $I_d$ dependence with respect to the temperature. Those represent two fundamental tests that any valid order parameter shall satisfy at transition points.

\begin{figure}[t]
\includegraphics[width=\columnwidth]{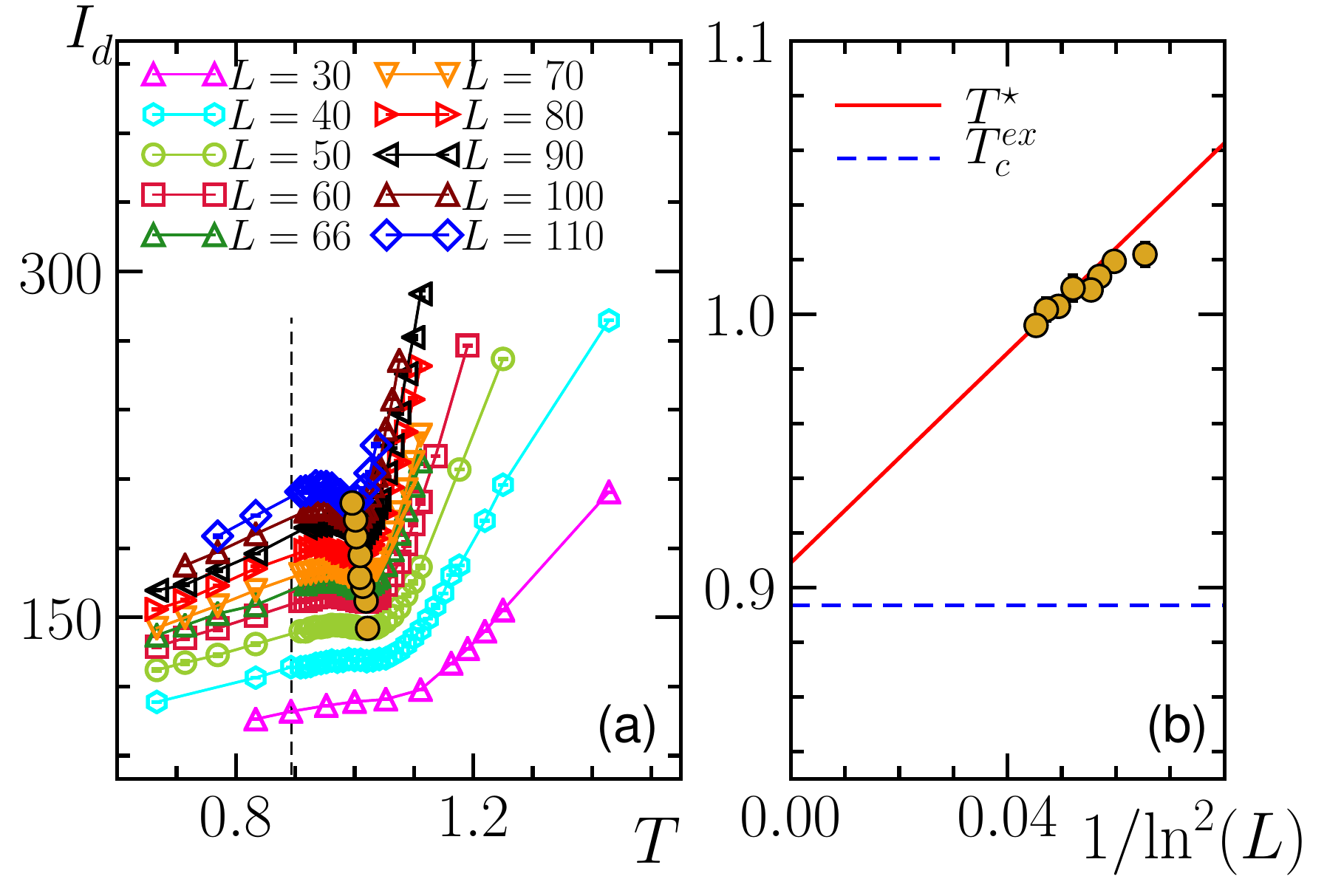}
\caption{\textit{Berezinskii-Kosterlitz-Thouless transition.} Panel (a) shows the temperature dependence of $I_d$ for different values of $L$. 
The dashed line indicates the value of the BKT critical temperature obtained in Ref. \cite{sandvik2013} using conventional methods,  $T_{BKT} = 0.8935(1)$. 
For each point, we harvested approximated $10$ instances of the data set and average the resulting estimates for the $I_d$.
The error bars are the standard deviation of such set of results. The scaling collapse obtained with these data sets is depicted in Fig.~\ref{fig1} (c).
Panel (b) shows the finite-size scaling of  $T^*$ based on Eq. \eqref{FSSxy}.
Fitting the results for $L = 80,90$, $100$ and $110$, we obtain $T_{BKT} =0.909\pm 0.015$.
In the text, we discuss how we obtain the local minimum of $I_d$, $T^*$.
We compute the $I_d$ of manifolds with $N_r = 5 \times 10^4$ configurations.}
\label{fig3}
\end{figure}

\begin{figure}[t]
\begin{center}
\centering
\includegraphics[width=\columnwidth]{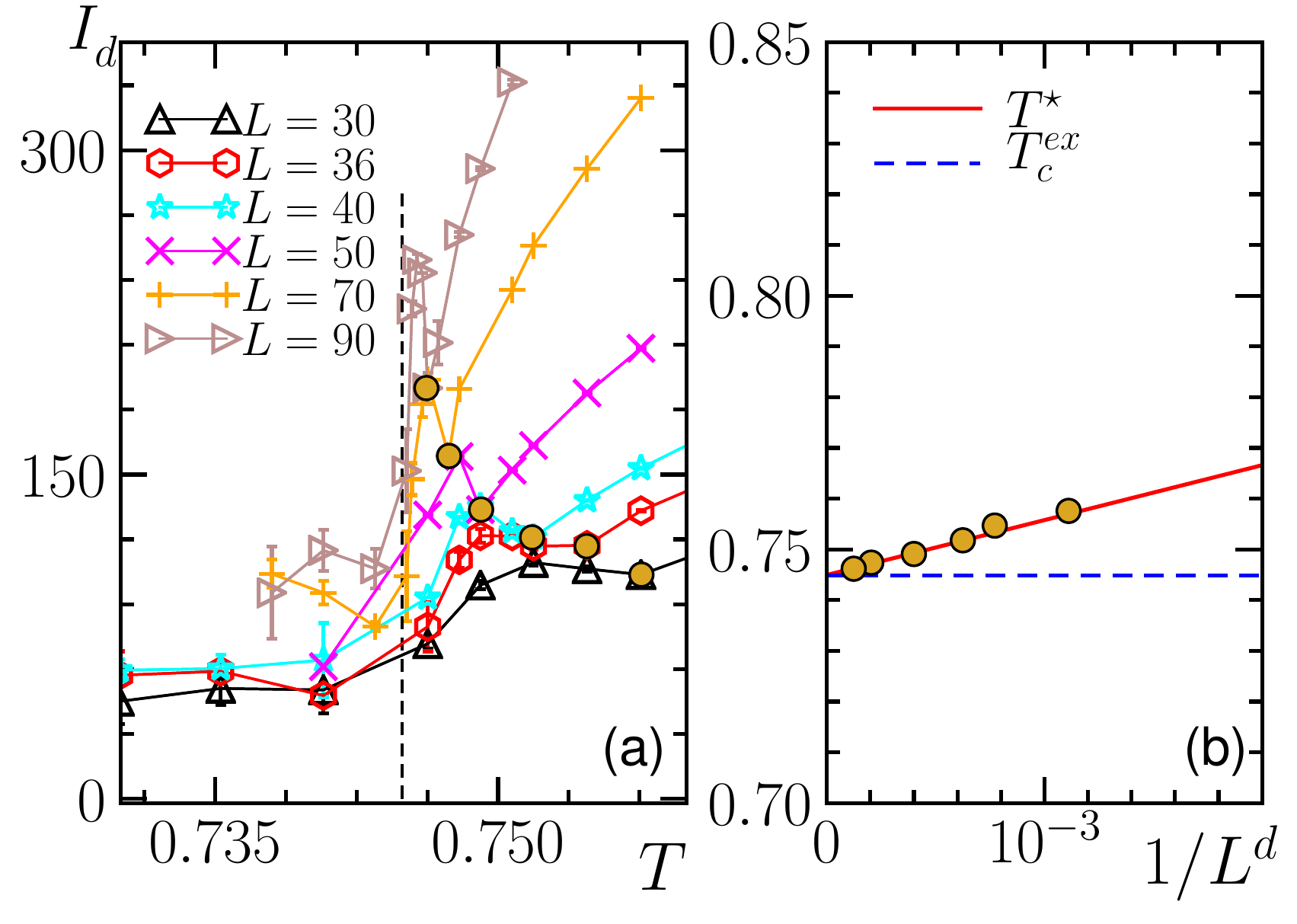}
\end{center}
\caption{\textit{First-order phase transition.}  Panel (a) shows $I_d$ as function of $T$ for the 8-states Potts model.
Panel (b) shows the finite size (FSS) scaling of $T^*$, where $d = 2$ (see text).}
\label{fig4} 
\end{figure}

\subsection{Berezinskii-Kosterlitz-Thouless (BKT) phase transition}

Unsupervised learning of phase transitions associated with the break of a symmetry, as discussed in the previous section, can also be performed with other unsupervised methods, such as PCA \cite{LeiWang2016,CeWang2017,Scalettar2017} and variational autoencoder (VAE) \cite{Wetzel2017}.
For example, the critical temperature of the Ising model can be obtained with an accuracy similar to the ones obtained here.
Furthermore, the latent parameters of the VAE can be used to learn the local order parameter associated to both discrete and continuous symmetry-breaking transitions \cite{Wetzel2017}.
Those methods are based on a dimension reduction, and thus, differ on a fundamental way from our unsupervised approach, which is based solely on the analysis of the $I_d$. As we discuss in this section, our approach can be extended to transitions that are characterized by non-local order parameters, such as the topological BKT phase transition, that are treated on the same footing of second order transitions.

The difficulties in learning the BKT transition from raw XY configurations occur in both supervised \cite{Melko2018} and unsupervised \cite{Scalettar2017} ML approaches. 
Recent progress based on diffusion maps \cite{Scheurer2019,Lidiak2020,nori2020} or topological data analysis\cite{Yoshihiko2020} have been made to solve this problem, typically considering problem-specific insights (such as the structure of topological excitations). These approaches have shown how considerable qualitative insight can be gathered on the nature of the BKT transition. However, it is presently unclear if the raw data structure corresponding to topological transitions can exhibit universal features, and if so, if unsupervised approaches can be used to detect the critical temperature with an accuracy that is
comparable with conventional methods (that typically rely on the a priori knowledge of the order parameter).

In Fig. \ref{fig3} (a), we show the temperature dependence of $I_d$ in the transition region. The intrinsic dimension clearly distinguishes
the low-$T$ regime, characterized by bound vortex-antivortex pairs, from the unbinding high-$T$ regime.
In the vicinity of the BKT critical point, $T_{BKT}$, the behavior of $I_d$ resembles the one observed for the second-order phase transitions,
i.e., $I_d$ exhibits a local minimum at $T^*(L)$ (observed for $L > 30$), which is a signature of the BKT transition. Note that the minimum is clearly visible already for lattices of order $L=50$; at these sizes, the spin stiffness is instead featuring a very smooth behavior, as considerably larger systems are required to appreciate a qualitative jump in the latter.

We consider the conventional FSS for BKT transition, 
$ I_d(T,L) = L^{\zeta} f(\xi(T)/L)$,  where 
the singular value of the correlation length diverge exponentially, i.e.,  $\xi \sim \exp{\left( a/\sqrt{T - T_c}  \right)}$.
In Fig. \eqref{fig1} (C), we show the universal data collapse for different values of $L$,
where  $a$, $T_{BKT}$ and $\zeta$ are treated as free-parameters in the collapse procedure; see the Appendix \ref{collapse}.
The value obtained,  $T_{BKT} = 0.92(1)$, is in good agreement with estimations of $T_{BKT}$ obtained on Ref. \cite{sandvik2010}.

A more accurate estimation of $T_{BKT}$ is based on the finite-size scaling of $T^{*}(L)$.
This approach relies on the computation of $T^*(L)$, which is performed  with the same procedure described in the previous section,
and the finite size scaling ansatz~\cite{sandvik2010}:
\begin{equation}
 T^*(L) - T_{BKT} \sim \frac{1}{\ln^2 L}.
 \label{FSSxy}
\end{equation}
As discussed before this procedure allows us to establish an error bar for the calculated $T_{BKT}$. 
We obtain  $T_{BKT} = 0.909 \pm 0.015$, that is compatible within error bars with Ref. \cite{sandvik2013} where simulations with up to $O(10^6)$ spins were carried out. For comparison, the best alternative method~\cite{Scheurer2019} utilizing unsupervised learning techniques reported relative errors of the order of 5\%.

Conventionally, $T_{BKT}$ is obtained with the aid of the so-called Nelson-Kosterlitz universal jump of the spin wave stiffness \cite{Nelson1977},
which allows to determine the finite-site critical temperature, $T_{BKT}^*(L)$. 
The FSS [Eq. \eqref{FSSxy}] is then used to determine BKT critical point at the thermodynamic limit.
Remarkably, here we observe that the intrinsic dimension of raw XY data sets exhibit a clear signature of the finite-site $T_{BKT}$,
even for moderate system sizes we have considered.

\subsection{First-order phase transitions}

Finally, we consider an example of  first-order phase transition (1PT): the $8$-state Potts model (8PM).
As is typical of  1PT the system exhibit a finite-size correlation length at $T_c$, $\xi_8 = 23.9$ \cite{Buddenoir1993}.
For $L > \xi_8$, the transition can be described by  trivial and generic critical exponents, e.g., 
$\nu = 1/d$, $d$ being the system dimension \cite{Fisher1982,Binder1984,sandvik2019}.
Furthermore, the finite-size shift of the critical temperature, $T_c(L)$, conventionally detected, for example, by the maximum value of the magnetic susceptibility, 
scales as~$T_c(L) - T_c \sim 1/L^d$.

Fig. \ref{fig4} shows that $I_d$ also exhibit a clear signature of the 1PT, featured by a peak at $T_c$ for $L \gg \xi_8$.
For $L \approx \xi_8$, the temperature dependence of $I_d$ resembles the one observed for 2PTs in Fig. \ref{fig2};
i.e., $I_d$ exhibit a local minimum at a temperature $T^*$.
Interesting, the  FSS of $T^{*}$ is in agreement with first-order transitions, see Fig. \ref{fig4} (b) \cite{Fisher1982,Binder1984}; 
the discrepancy of the calculated $T_c=0.7448(1)$ with the exact value is less the $0.05\%$

\section{Discussion}

Our results so far support the fact that, in the vicinity of a phase transition, the intrinsic dimension displays universal behavior at both first, second-order and BKT transition, and works as a order parameter signalling a transition between different data structures in configuration space. Within this framework, the position of the transition is always identified with the scaling of the minimum of the intrinsic dimension. 

In continuous phase transitions, collective behavior is captured by only a handful of parameters. This suggests that the amount of information required to describe the system is parametrically simpler at the critical point when compared to its vicinity, as the latter region requires additional information on the operators required to perturb away from criticality. This emergent simplicity may have several consequences at the data structure level. The most direct consequence is that one expects a simplified data structure to be described by a minimum of the intrinsic dimension at the transition point. This is exactly what we have observed at both second-order and BKT transitions. We note that this expectation is not related to the number of states sampled by the partition function (this number is, in our case, fixed by $N_r$ are configurations are never repeated).
The discussion of how our results change with $N_r$ is reported in Appendix \ref{scalingNr}.

For first-order transitions, the above reasoning is not applicable as it relies on universal behavior, and thus the existence of a continuum limit. In these cases, one expects  that the data space in the vicinity the transition point shall feature two separate regions, each of them composed of states representing the two phases meeting at $T_c$. Exactly at the transition point, one expects an abrupt change in the data structure: indeed, the MC sampling will access a large number of configurations corresponding to both phases (in analogy to metastability), and thus display a sharp increase (see Fig.~\ref{fig4} (a)). Approaching the transition point from the disordered phase will feature instead of a minimum, that scales to the transition point. 

The arguments above serve as a qualitative guideline behind the basic picture we put forward: the simplified field theory description applicable at transition points reflects directly into the data structure of the problem. We now provide a data-driven discussion in support of this picture, that specifically emphasize the connection between the data set and correlations in the system via the (generic) definition of distance we employ. For the sake of concreteness, we first elaborate on the presence of distinctive features in the vicinity of $T_c$, and then connect to universal scaling.

\subsection{Why the $I_d$ exhibits a singular behavior in the vicinity of $T_c$?}

\begin{figure}[]
\begin{center}
\centering
\includegraphics[width=\columnwidth]{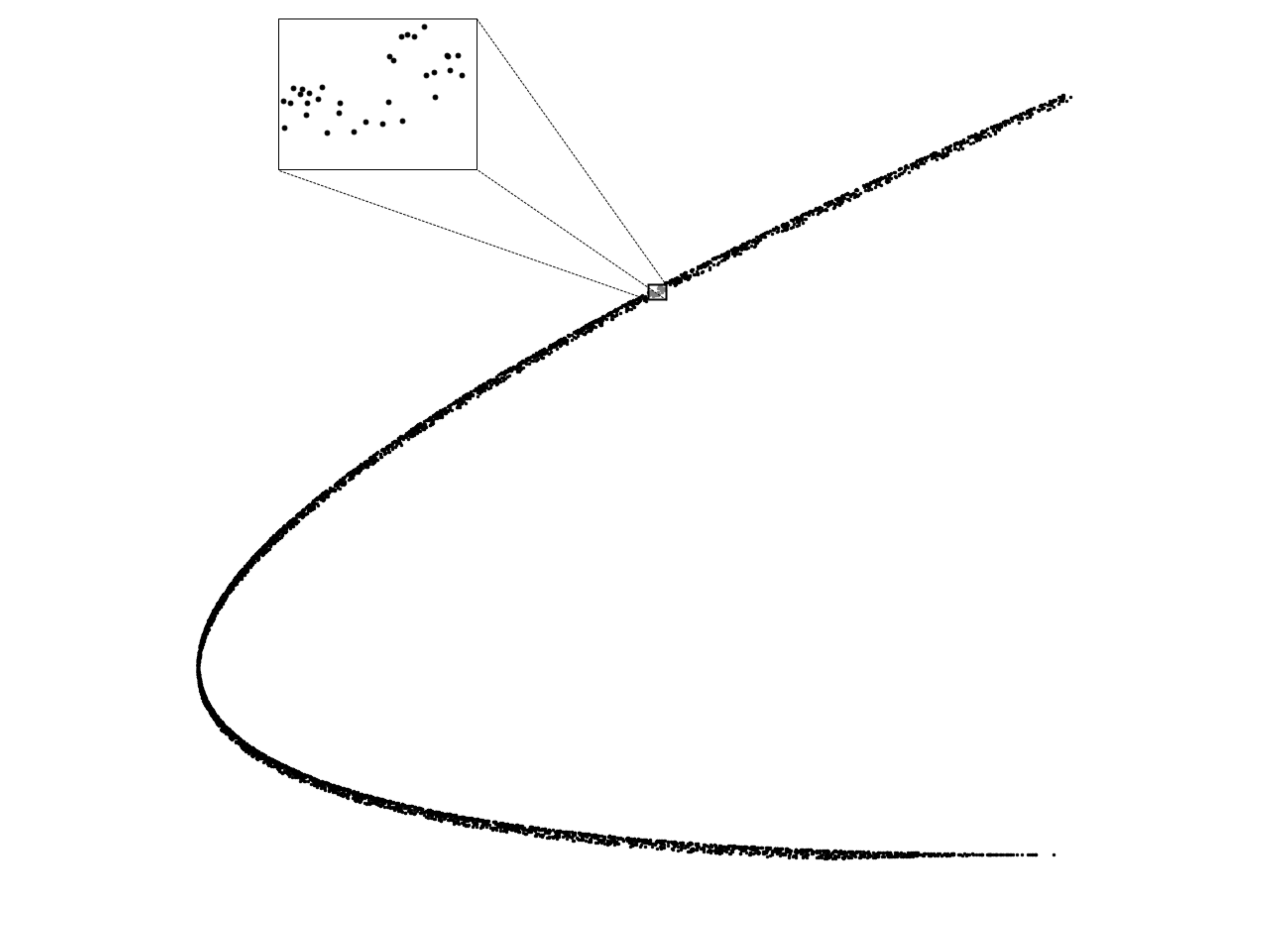}
\caption{\textit{Scale dependence of ID}. The data set shown presents an $I_d=1$ or $I_d=2$ depending on the scale that is considered.   }
\label{figscale} 
\end{center}
\end{figure}

The $I_d$ is a \textit{scale} dependent quantity~\cite{Rozza2015}. This can be intuitively understood by looking at example depicted in Fig.~\ref{figscale}, where an approximately one dimensional object appears as two dimensional when looking at a different scale by zooming.
The scale of the data set, as estimated with two-NN, is fixed by $N_r$ for a given $T$, since it fixes the actual meaning of first and second nearest neighbors \cite{Laio2017}; here we always consider $N_r = 5 \times 10^4$.
In the following, we will show how changes in the scale of the data (configuration) space appear when there is a phase transition, leading to the emergence of features in the $I_d$. 

\begin{figure}[]
\begin{center}
\centering
{\centering\resizebox*{9cm}{!}{\includegraphics*{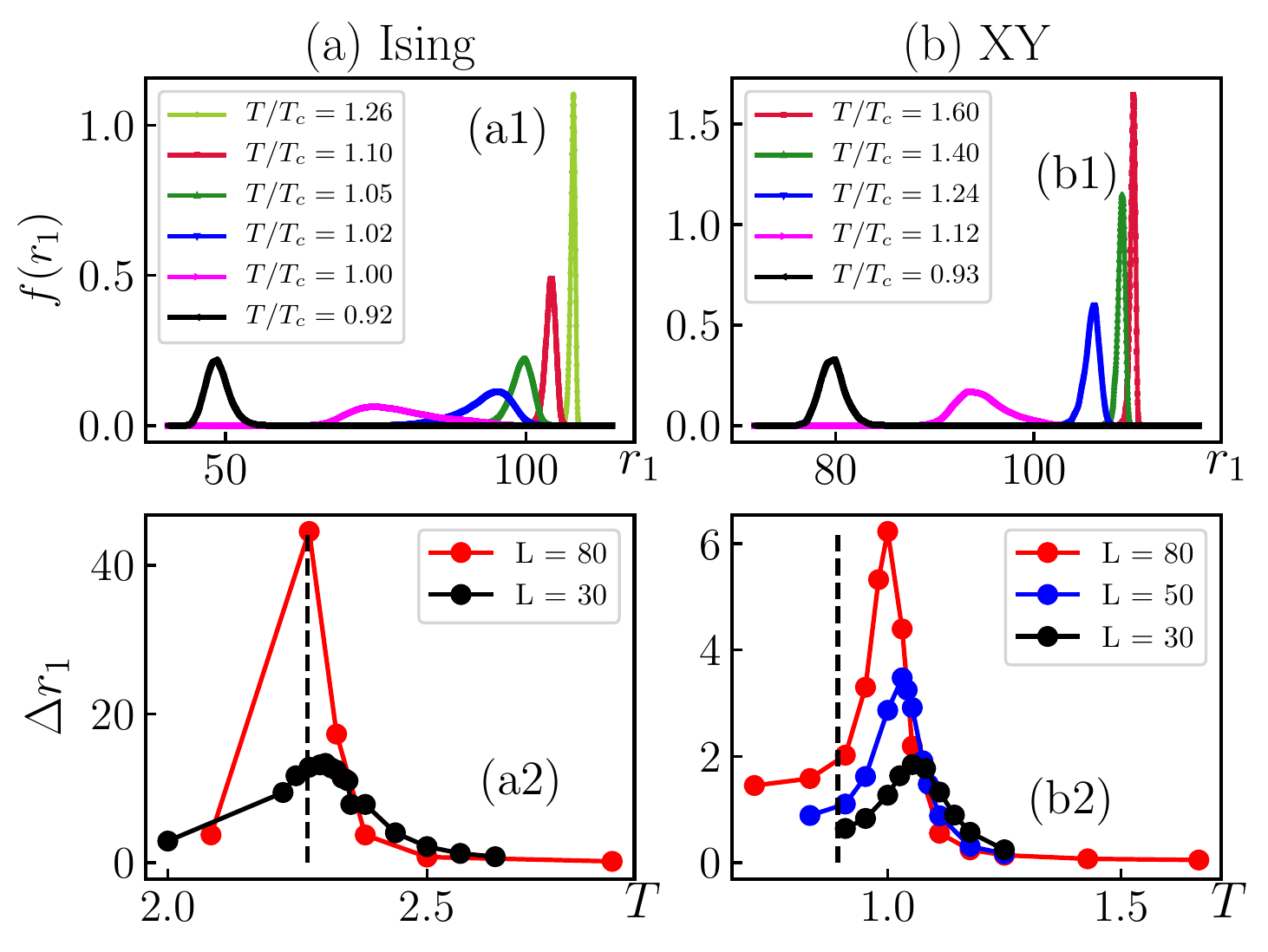}}}
\end{center}
\caption{\textit{Statistics of first nearest-neighbor distances, $r_1$.} In panels (a1) and (b1) we show the  Kernel density estimation of the probability density function of $r_1$ for the Ising and XY data sets, respectively.
All the results have $L = 80$.
Panels (b1) and (b2) show the temperature dependence of the variance associated to the distribution functions show in (a) for different system sizes. 
Results for the second nearest-neighbor distances, $r_2$, are qualitatively the same.}
\label{fig7} 
\end{figure}

A first test is to check that these changes in the scale effectively occur. To this end, we analyze the statistics of $r_1$ and $r_2$.
For example, the distribution function of the first-neighbor distances, $f(r_1)$, changes for both Ising and BKT critical points, see Figs. \ref{fig7} (a1) and (b1).
The position of the peak of $f(r_1)$ sharply decreases as one crosses the transition, and the variance associated to $f(r_1)$, $\Delta r_1$, has a peak close to the transition, see Fig.~\ref{fig7} (a2) and (b2).
Interestingly, our results indicate that the quantity  $\Delta r_1$ also exhibit universal scaling behavior at Ising and BKT critical points. Moreover, the results for both Ising and XY data sets are qualitatively the same, highlighting that data-wise, symmetry-breaking and topological transitions are treated on the same footing. 
However, it is important to stress that, contrary to what happens in the case of $I_d$, the peak in $\Delta r_1$ is not expected to present features when the data sets are not homogeneous in density, since the relevant distances will be also affected due to these inhomogeneities (see ref.~\citenum{Laio2017} for a further discussion on why two-NN is only mildly affected by this problem). In this sense, the intrinsic dimension, being solely sensitive to changes of scale and not local density features, provides a considerably more reliable probe for phase transitions.

\begin{figure}[]
\begin{center}
\centering
\includegraphics[width=\columnwidth]{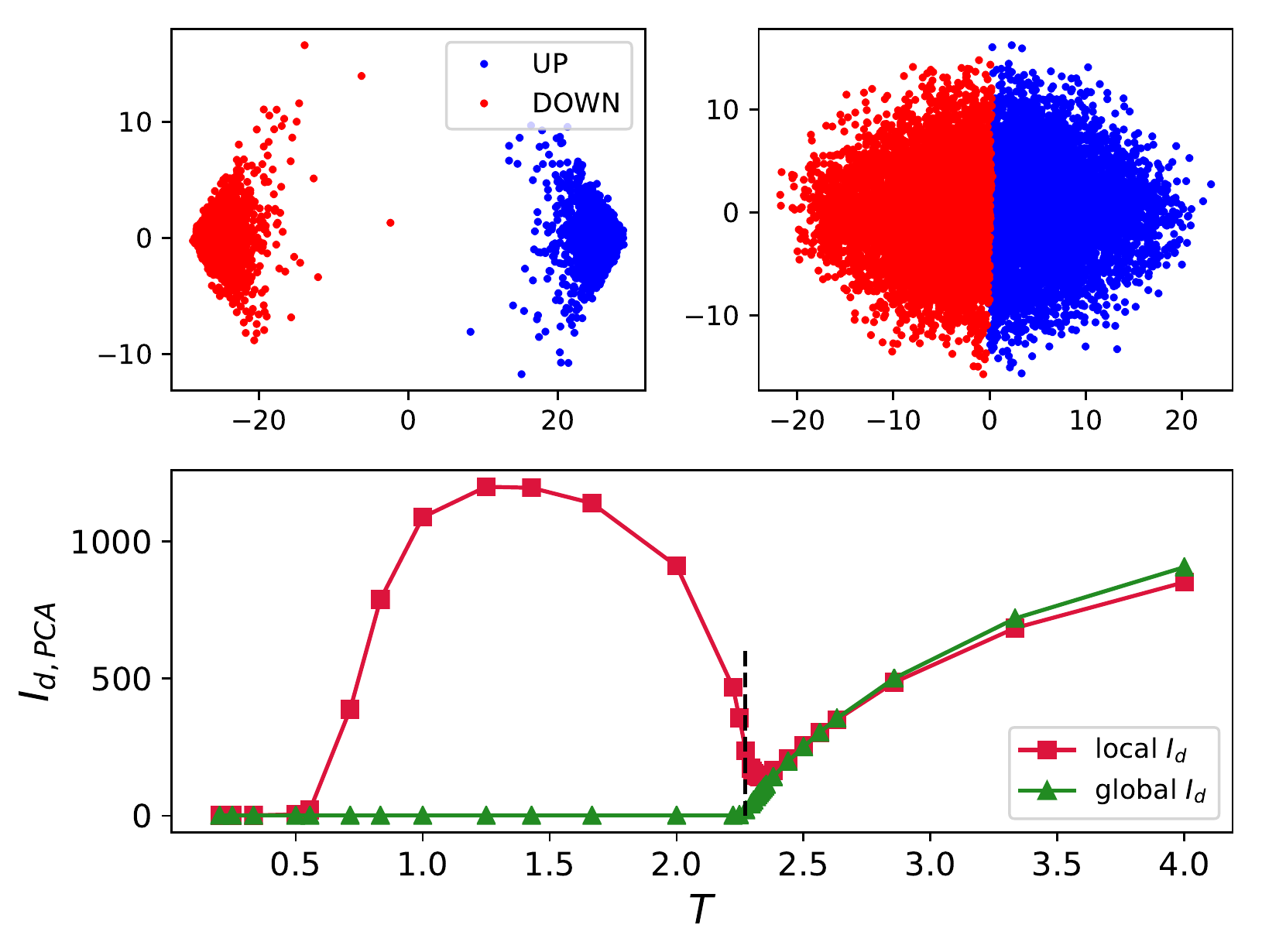}
\end{center}
\caption{\textit{$I_d$ obtained with PCA.}  Panel (a1) and (a2) show the projection of the Ising data set in the two leading principal components for $ T/T_c \approx 0.94$ and $T/T_c \approx 1.10$, respectively.
Configurations with total magnetization $M > 0$ are represented by the blue points, while  ones with $M \le 0$ by the red points.
Panel (b) shows the PCA estimation of the $I_d$ considering the full Ising data set (global $I_d$) and the data set generated by configurations with total magnetization $M > 0$ (``local'' $I_d$). 
For all the results, $L = 60$. }
\label{fig5} 
\end{figure}

\begin{figure}[]
\begin{center}
\centering
{\centering\resizebox*{8cm}{!}{\includegraphics*{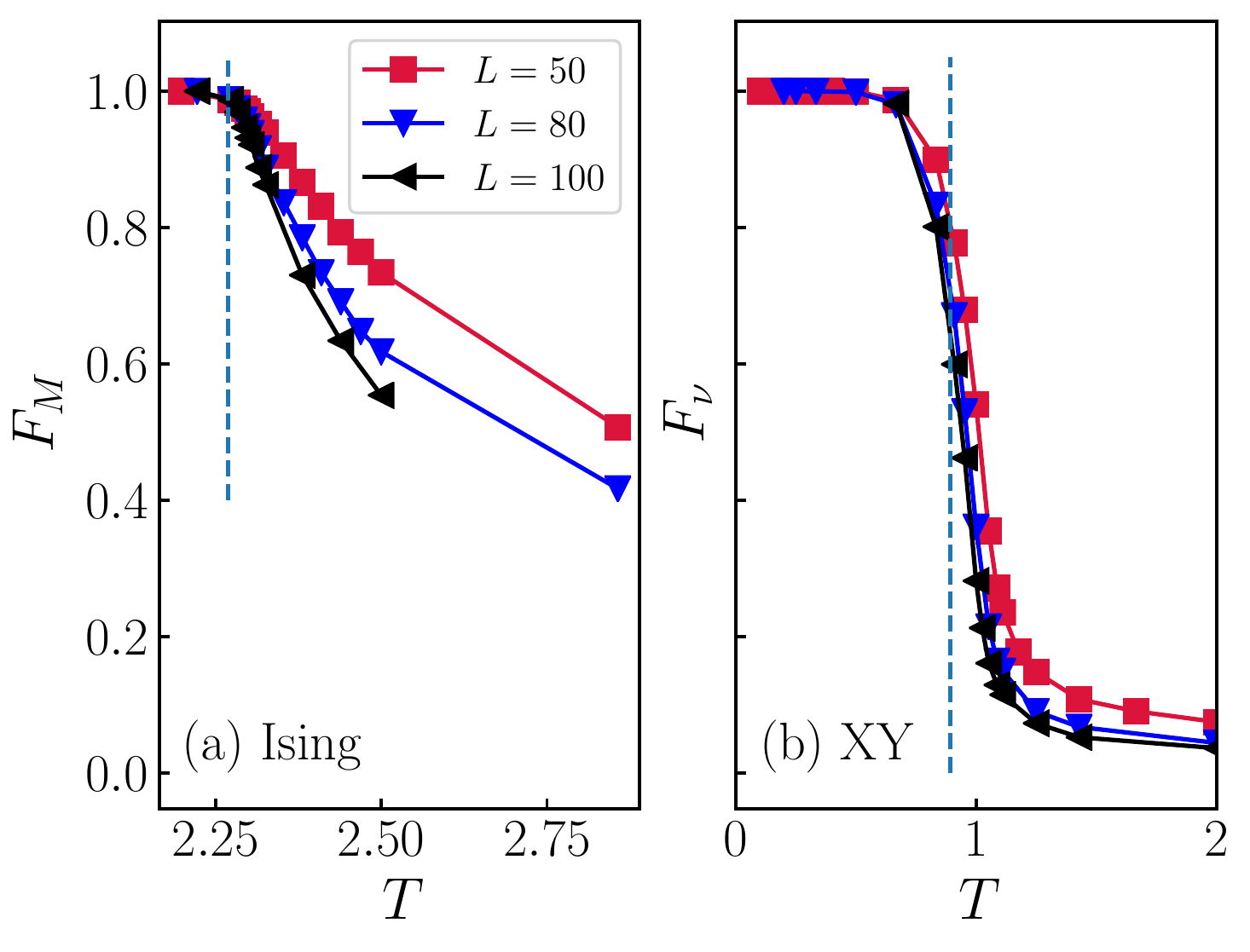}}}
\end{center}
\caption{\textit{Connectivity between neighboring points.} Panel (a) shows the fraction of points in the Ising data set whose first two nearest-neighbors have the same magnetization sign. 
Similarly, in panel (b), we  show the fraction of points in the XY data set whose first two nearest-neighbors have the same winding number (see text).}
\label{fig6} 
\end{figure}

In order to understand the underlying cause to this change of scale, we first focus on discrete symmetry-breaking transitions.
In those cases, PCA provides an understanding of the data structure emerging at critical points \cite{LeiWang2016,Scalettar2017}.
For instance, the Ising data set features clusters characterized by configurations with positive,  $+M$,  and negative, $-M$, total magnetization for $T < T_c$.
In contrast, a single cluster is formed for $T > T_c$, see Figs. \ref{fig5} (a1) and (a2).
This clustering structure allow us to understand the connectivity between neighboring configurations in the Ising data set.
For $T > T_c$, the magnetization of neighbors is completely random.
In contrast, configurations connect to first and second neighbors with the same magnetization sign for $T < T_c$; see Fig. \ref{fig6} (a).
Equivalent reasoning based on PCA is applicable to the  Potts data sets.

To illustrate how the locality (and the the connectivity between neighboring configurations) affects the behavior of the $I_d$, 
we consider two estimates of the $I_d$ provided by PCA. 
In the first case, we employ all the configurations of the Ising data set, $I_{d,PCA}$ (global) while, in the second, we consider just configurations with M > 0, $I_{d,PCA}$ (local); see the Appendix \ref{distances} for more details.
The latter quantity provides a local estimate (within the range scale of a single cluster) for $T < T_c$, which is analogous to the local measure of the $I_d$ provided by the TWO-NN method. 
As shown in Fig. 7 (b), the global $I_{d,PCA}$ sharply goes to 1 below $T_c$, while the local $I_{d,PCA}$ exhibits the same non-monotonic behavior close to $T_c$ observed in Fig. \ref{fig2}. 
This result highlights that the locality of the $I_d$ is the crucial element to understand its non-monotonic behavior close to $T_c$.

The connectivity between neighboring configurations is also related to the physical properties of the BKT transition.
In this case, the most suitable quantity to characterize configurations and the corresponding clustering structure in phase space is the winding number, $w$ (see Appendix \ref{winding}), as excitations have topological (global) nature~\cite{Scheurer2019}.
Above the BKT transition, vortex-antivortex pairs are unbounded. Due to this, MC simulation samples configurations with different $w$.
By  contrast, in the quasi-long-range-order regime ($T < T_{BKT}$), most of the configurations have  $w = (0,0)$. 
This feature of the BKT transition affects the connectivity between neighboring configurations.
In particular, the fraction of configurations  whose first two neighbors are connected to points with the same $w$, $F_{w}$, is negligible for $T > T_{BKT}$, but it is equal to $1$ in the topological phase. This is illustrated in Fig. \ref{fig6} (b).
Thus, for $T < T_{BKT}$, the $I_d$ is a property of the manifold exclusively formed by configurations connected to neighbors with the same winding number.

In a nutshell, the underlying cause for the sensitivity of the $I_d$ to phase transitions is that, both the symmetry-breaking and topological transitions, affect the neighboring configurations' connectivity.
The key aspect is that nearest-neighbors configurations have identical physical properties (order parameter and winding number) when the system is in the ordered phases (symmetry-broken or quasi-long-range-ordered). 
By contrast, in the disordered phase, the first and second neighbors' physical properties are entirely random.
At the data structure level, the phase transition represents a change of scale between those regimes.

\subsection{Why $I_d$ exhibit universal scaling behavior?}

Based on the fact that $I_d$ shows a characteristic minimum feature in the vicinity of phase transitions at $T_c(L)$, we now provide an argument in support of universal scaling of the latter temperature against system size. The key aspect of our argument is that the distances $r_1$ and $r_2$ are related with many-body correlation functions in the system, computed at equilibrium.

We analyze the curve $\ln (1-P_i)$ versus $\ln \mu_i$ close to the origin. From Eq.~\eqref{Id}, the slope of this curve is proportional to $I_d$. The curve starts at the origin; we assume that its slope can be correctly determined by sampling the first point of the curve several times (e.g., by sampling several independent Markov chains); this seems very well satisfied based on our earlier numerical observations (fluctuations and deviations from linear behavior typically appear only for very large values of $\mu$).  Within this assumption, one obtains the following estimate for $I_d$:
\begin{equation}\label{dist}
I_d = - \frac{\ln(1-1/N_r)}{\ln [r_2(1)] - \ln [r_1(1)]}.
\end{equation}
We post-pone to the end of section an alternative justification for such a scaling behavior. 
From now on, we specifically consider the Euclidian distance function [see, Eq.\eqref{diseuclidian}]; using the hamming distance will not affect the substance of our reasoning, but will change some of its details.

For the sake of simplicity, we can assume that the reference configuration $i=1$ corresponds to the lowest energy state. This second assumption relies on the fact that such state is the one that has a higher probability of being sampled at any temperature, and, at least at sufficiently low temperatures, it is very likely to be the state with the lowest value of $\mu$, as low-lying excitations do typically differ from the lowest energy states by a low amount of spin flips (representative of spin waves), when compared to the average distance between states. 
Within this approximation,  we can fix the coordinates of the reference configuration: $s_{j}=s\quad \forall j$ for the Ising data set and $\theta_{j}= \theta \quad \forall j$ for the XY data set.

\begin{figure}[]
\begin{center}
\centering
{\centering\resizebox*{8.5cm}{!}{\includegraphics*{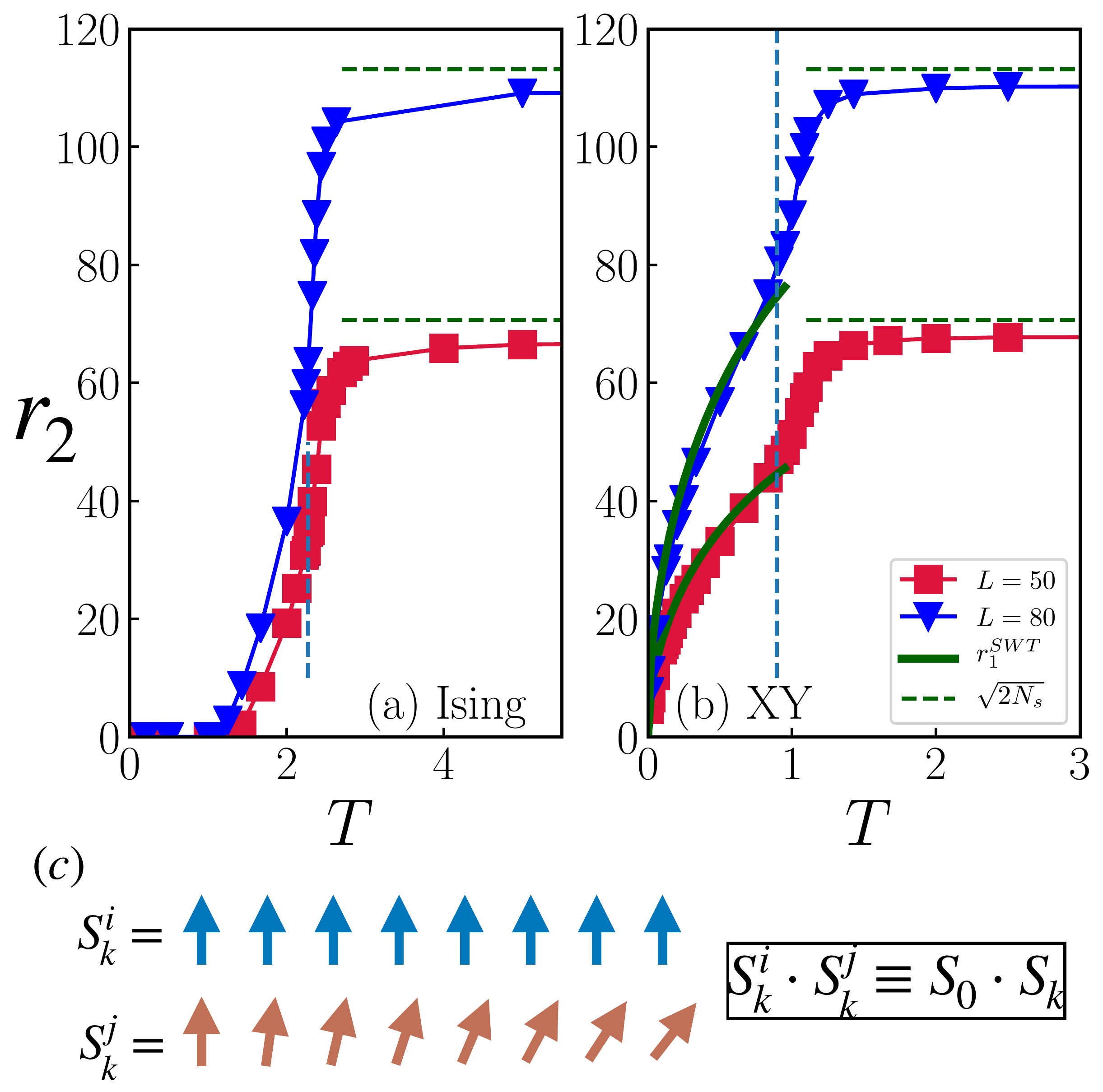}}}
\end{center}
\caption{\textit{Correlation functions and distance between neighboring configurations.} Temperature dependence of the smallest $r_2$ (this distance is representative of the lowest value of $\mu$, see text) for the (a) Ising and (b) XY models.
The lines (full and dashed) represent the predictions for $r_2$ based on Eq.\,\eqref{eqr1} and the corresponding  expressions of the asymptotic formula for the correlation functions in the high and low temperatures regimes (see text). The panel (c) illustrate the basic assumption behind Eq. \eqref{eqr1}, i.e, correlation between configurations are equivalent to correlation functions(see text).
}
\label{fig9} 
\end{figure}

We can now proceed and analyze the denominator of Eq.~\eqref{dist}. We define:
\begin{equation}
    \alpha_{0f} = \sum_{j=1}^{N_s} S_0 S_{j,f},
\end{equation}
where $S_0$ represent the coordinates of the reference configuration; see Fig. \ref{fig9} (c).
Thus, the distance between two configurations reads:
\begin{equation}
    r_f(1) =\sqrt{2N_s} \sqrt{1 - \frac{\alpha_{0f}}{N_s}}
    \label{eqr1}
\end{equation}
We then get:
\begin{equation}
    \ln [r_2(1)] - \ln [r_1(1)] = \frac{\ln (1 - \frac{\alpha_{01}}{N_s}) - \ln (1 - \frac{ \alpha_{02}}{N_s})}{2}.
\end{equation}
For $T \gg T_c$, the coordinates of neighboring configurations are expected to be completely random compared to $S_0$. Thus,  it is reasonable to expect that $\alpha_{0f}\ll N_s$  (we will come back to this point below). Analyzing the transition from the disordered phase, we can expand the logarithms up to second order in $\alpha_{0f}/ N_s$ and get:
\begin{equation}
    \ln [r_2(1)] - \ln [r_1(1)] = \mathcal{F}_{2} + \mathcal{F}_{4} + ...
\end{equation}
where the function $ \mathcal{F}_{p}$ contains all $p$-spin correlation functions $S_{j_1,f}S_{j_2,f}...S_{j_p,f}$ taken over the single states 1 and 2. In principle, one shall also retain other orders: in fact, the difference of the two distances depends parametrically on arbitrary body correlation functions. 

Now, we make a third assumption, that is, that the correlations contained in $ \mathcal{F}_{p}$ can be replaced by the corresponding thermal averages. The rationale behind this is that, based on our first assumption above, we are actually considering the states that have the highest weight in the partition function, so the ones that contribute the most to the computation of the correlation function. Here, temperature plays a clear physical role: higher temperature let us sample states that are (on average) at a larger distance from the lowest energy state when compare to lower temperatures. One can reformulate the above as follows. For any given Markov chain, we have a given $ \mathcal{F}_{p}^{(k)}$, which depends on correlations on a single pair of configurations. Then, average over the various Markov chains gives us an averaged value, that depends on the average of correlations over the various configurations. This last formulation is closer to the numerical recipe that we utilize to estimate $I_d$, where, in fact, we obtain the latter from averaging the $I_d$ resulting from several distinct simulations.

Now, since we are dealing with thermal averages, we can recall the finite-size scaling hypothesis. This hypothesis tells us that, if a quantity develops a singular behavior at the transition point (not necessarily a divergence), the temperature corresponding to such a feature $T_{\text{feat}}$ shall be shifted according to finite-size scaling (FSS) theory as (for second order phase transitions):
\begin{equation}
    (T_{\text{feat}}-T) \propto \frac{1}{L^\nu } 
\end{equation}
We are thus in a position to make a statement: if any of the arbitrary body correlation functions contained in the definition of our distance displays singular behavior at the transition point, those will dictate the scaling of the position of the minimum of $I_d$ according to FSS, and reveal us the critical exponent $\nu$ (or, in case of BKT, they will be consistent with the logarithmic scaling expected there). The behavior of all other correlations is not expected to affect this scaling behavior at all, as those are not displaying any non-singular feature by definition.

We note that our reasoning and the validity of its assumptions can be {\it a posteriori} verified by noticing that it implies that features in the distribution of the distances $r_1$ will also be related to critical behavior. 
In particular, we consider the pivotal assumption that the correlation contained in Eq.~\eqref{eqr1} can be replaced by the corresponding thermal averages, and compare the predictions for $r_f(1)$ with our numerical results~\cite{ft1}.
Fig. \ref{fig9} shows this comparison for both the Ising and XY models.
(i) In the disordered phases, $\alpha_{0,f} \ll 1$, due to the exponential decay of the correlations, and  thus $r_f \approx \sqrt{2N_s}$. (ii) On the other hand, in the symmetry- broken phases, $\alpha_{0,f} \approx O(N_s)$, given the long-range nature of the correlations; which implies that $r_f \ll 1$. (iii) Finally, in XY model's critical phase, the temperature dependence of the correlations is given by $S_0 \cdot S_{j,f} \sim |j|^{-T/2\pi}$ (where |j| represents the spatial distance from a reference site). By computing the corresponding $\alpha_{0,f}$, one can obtain the temperature dependence of $r_f$. The numerical results display very good agreement with our predictions; see Fig. \ref{fig9} (a) and (b). It is worth noting that, while our argument justified critical scaling for $T_c(L)$ and does not justify the full collapse scaling observed for $I_d$, this is still directly informative about both critical temperature, and the critical exponent $\nu$. 

Before ending the section, we present a different approach to determining the dependence between $I_d$ and the smallest value of $\mu$ as per Eq.~\eqref{Id}. An alternative way to qualitatively estimate $I_d$ from a data distributed according to Eq.~\eqref{eq:fmu} is to apply the maximum likelihood criterion. Utilizing the commonly used log-likelihood function $\ell_k = \log[f(\mu_k)]$, one obtains that, for data sets where $I_d\gg1$, one has $I_D\simeq N_r / \ln (\mu_1)$. The scaling with $N_r$ is different with respect to Eq.~\eqref{dist}: this is not unexpected due to the fact that (1) we are considering sampling of few configurations in the previous approximation, and (2) maximum likelihood does not necessarily capture the correct scaling with the number of points in the set (as one may expect many of those do not contribute to the determination of the minimum).  Nevertheless, this difference is irrelevant for the sake of our argument above, as we are not immediately interested in the $N_r$ scaling. What is important is that maximum likelihood returns exactly the same functional dependence on $\mu_1$, thus providing a data driven justification of the first assumption presented above.

\section{Conclusions}

We have shown that phase transitions can be learned  through a single property of raw data sets of configurations - the intrinsic dimension - without any need to perform dimensional reduction.
The key observation made here is that, in analogy to physical observables, the intrinsic dimension exhibits universal scaling behavior close to different classes of transitions: first-, second-order, and Berezinskii-Kosterlitz-Thouless (BKT). 
This indicates how the intrinsic dimension, in the vicinity of critical points, behaves as an order parameter in data space, showing how the latter undergoes a structural transition that parallels the phase transition identified by conventional order parameters.

At the practical level, we have shown that the finite-size analysis of intrinsic dimension allows not just to  detect,
but also to characterize critical points in an unsupervised manner. 
In particular, we have shown that the intrinsic dimension allows one to estimate transition temperatures and (critical) exponents of both first- and second-order transitions with accuracies ranging from 1\% to 0.1\% at very modest system sizes. In addition, the method is equally applicable to topological transitions, where we have demonstrated an accurate (with $1\%$ of confidence) estimation of the BKT topological transition competitive with more traditional methods at the same system sizes. This latter result suggests that the lack of any dimensional reduction allows retaining topological information in the vicinity of the phase transition, which may instead be lost otherwise \cite{Scalettar2017,Scheurer2019}.

A fundamental aspect of our approach is that it is based on a $I_d$-estimation method
suitable to learn complex manifolds, such as the twisted XY manifold emerging at the BKT critical point.
The results demonstrate the potential of state-of-the-art $I_d$-estimators \cite{Laio2017,erba2019intrinsic} methods to tackle  many-body problems, and motivates an even stronger methodological connection between data mining techniques, and many-body physics. We also note that, in comparison with previous applications in other fields \cite{Facco2019,ansuini2019intrinsic,rodriguez2018}, the values of the intrinsic dimension reported here are considerably larger. For future applications, like combining our analysis with clustering methods that do not rely on dimension reduction\cite{d2018automatic}, it may be interesting to develop novel estimators that focus on large values of $I_d$, potentially trading absolute accuracy with numerical efficiency (in the spirit of ref. \citenum{erba2019intrinsic}). Another interesting question to address in the future is to assess the possibility of data lying in submanifolds with different $I_d$\cite{allegra2020data} and how it affects to our method. This is a plausible scenario in cases with co-existence of phases.

Some of the methods presented here may be applied to quantum mechanical objects, such as quantum partition functions, density matrices, and wave functions. It is an open challenge to determine whether the data mining of quantum objects can provide an informative perspective on the latter, such as, e.g., accessing entanglement or other more challenging forms of quantum correlations. Finally, while we focused on configuration generated by Monte Carlo sampling, our approach is equally applicable to experimentally generated data; it may be interesting to apply it to  settings where raw data configurations are available, such as, e.g., quantum gas microscope experiments~\cite{doi:10.1142/9789814667746_0004,Scalettar2020,Annabelle2019}.

\section{Acknowledgements} 
We acknowledge useful discussions with R. Ben Ali Zinati, R. Fazio, A. Laio and R. T. Scalettar. The work of TMS, XT and MD is partly supported by the ERC under grant number 758329 (AGEnTh), by the Quantera programme QTFLAG, and has received funding from the European Union's Horizon 2020 research and innovation programme under grant agreement No 817482. This work has been carried out within the activities of TQT. 
TMS and XT acknowledge computing resources at Cineca Supercomputing Centre through the Italian SuperComputing Resource Allocation via the ISCRA grants ICT20\_CMSP and MLforPT.

\appendix

\section{The TWO-NN method  and comparison with principal component analysis (PCA)}
\label{distances}

In this section, we provide more details about the two-NN method.
As described in Ref. \cite{Laio2017}, the intrinsic dimension, $I_d$, can be obtained through the following steps: 
\begin{itemize}
 \item[1.]  For each point $i$ of the data set ($i = 1,2,...,N_r$), compute its first- and second-nearest neighbor, $r_1(i)$, $r_2(i)$, respectively.
 \item[2.]  For each point $i$, compute the ratio $\mu_i = r_2(i)/r_1(i)$.
 \item[3.]  The empirical cumulate is defined as $P^\textup{emp}(\mu) = i/N_r$, while the values of $\mu_i$ are sorted in an ascending order  through a permutation, i.e., ($\mu_1, \mu_2, ... \mu_{N_r}$),
 where $\mu_i < \mu_j$, for $i < j$.
 \item[4.]  Finally, the resulting ${S = \{ (\ln(\mu), - \ln\left[ 1 - P^\textup{emp}(\mu) \right]   \}}$ are fitted with a straight line passing through the origin.
 The slope of this line is equal to $I_d$ (see Eq.\eqref{Id}).
\end{itemize}

\begin{figure}[t]
{\centering\resizebox*{8.8cm}{!}{\includegraphics*{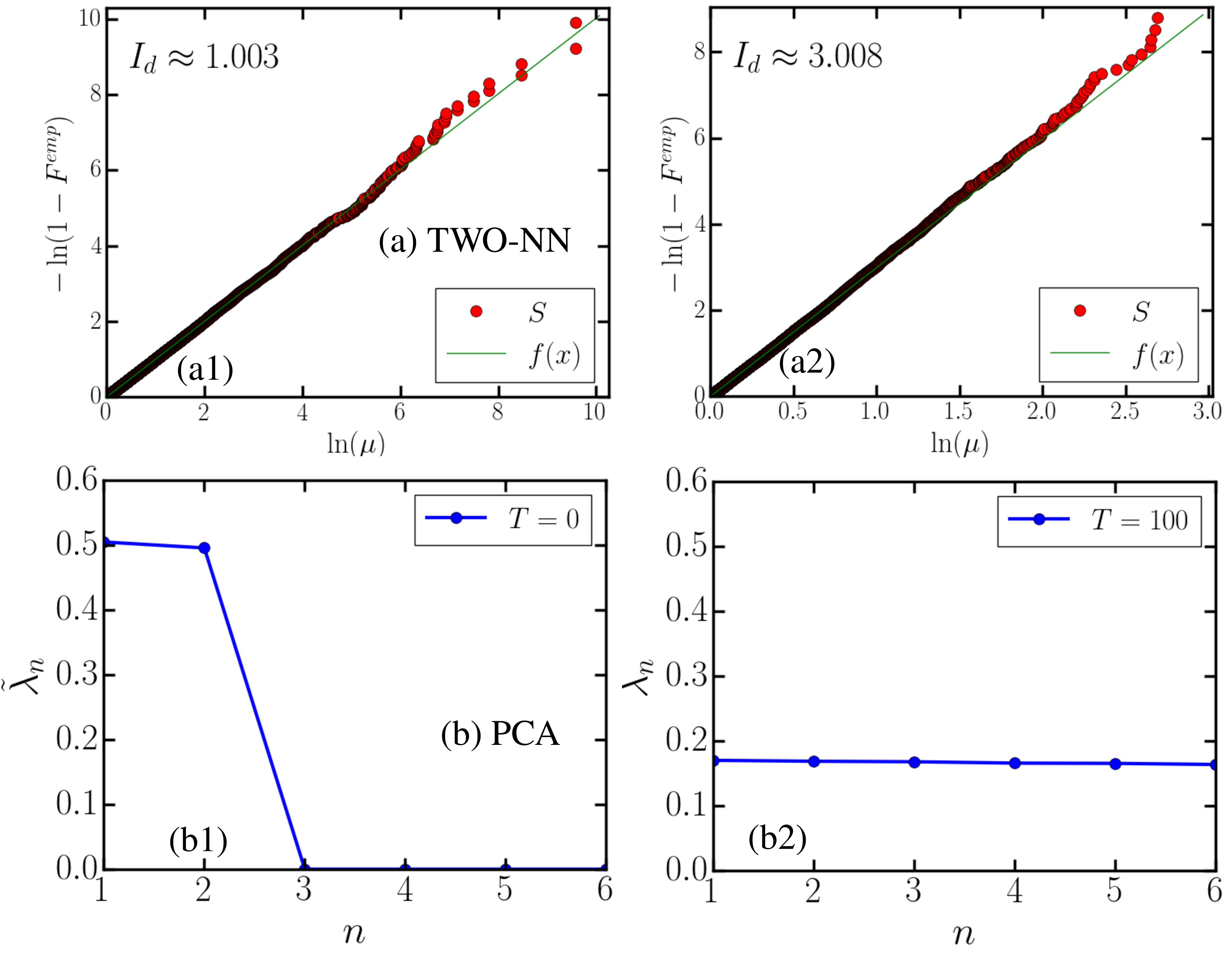}}}
\quad
\caption{\textit{3-site XY model.} Panels (a) show results of the TWO-NN method: fitting of the data points $S$ for (a1) $T = 0$ and (a2) $T = 100$. The data set have $N_r = 10^3$ configurations.
We obtain (a1) $I_d = 1$ and (a2) $I_d = 3$.
Panels (b) show results of the PCA method: normalized eigenvalues of the covariance matrix, $\tilde{\lambda_n}$, obtained from the raw XY configurations. Here we use the same notation of ref. \cite{Scalettar2017}
For $n > I_d^{PCA}$, $\tilde{\lambda_n} \to 0$. PCA predicts (b1) $I_d^{PCA} = 2$ and (b2) $I_d^{PCA} = 6$, which is not in agreement with the exact results (see text)}
\label{fig1SM}
\end{figure}

\begin{figure}[t]
{\centering\resizebox*{8.8cm}{!}{\includegraphics*{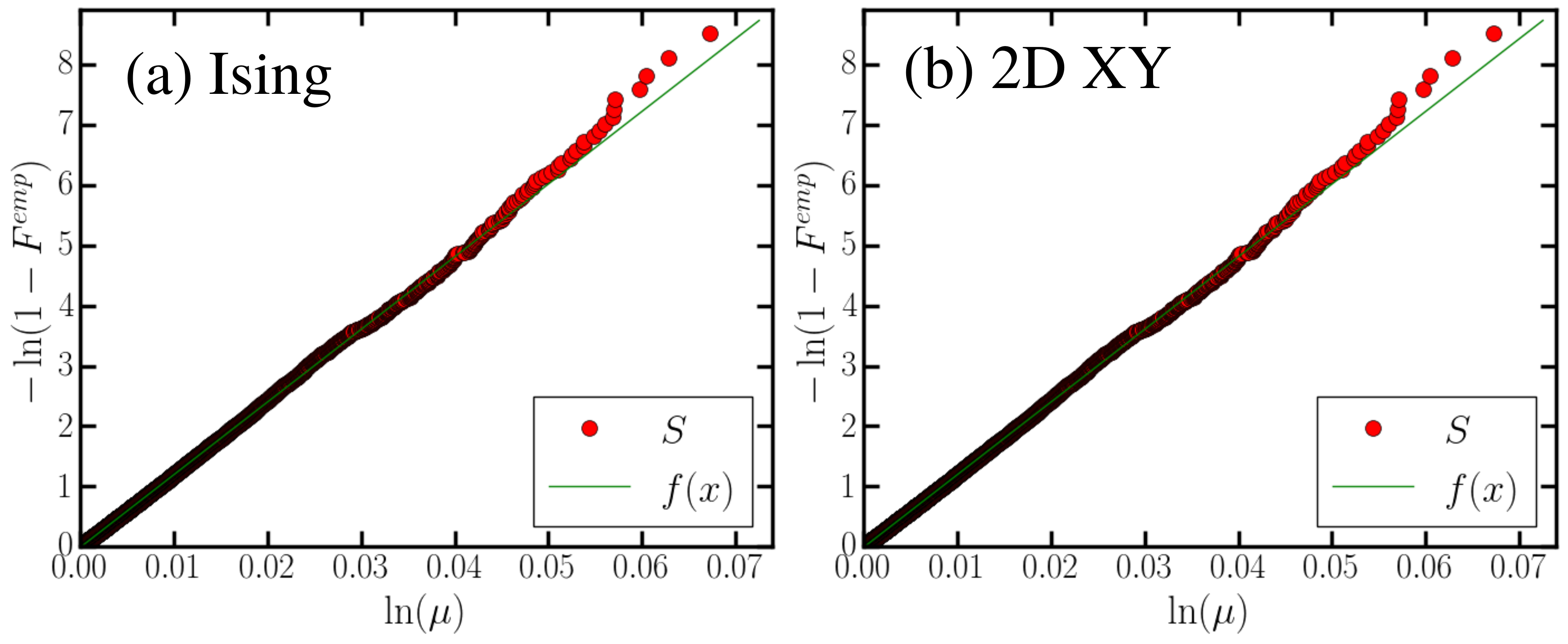}}}
\quad
\caption{ Results of the TWO-NN method: fitting of the data points $S$ for (a) Ising and (b) 2D XY thermal data sets generated close to the critical point (in (a) $T \approx 2.2$, while in (b) $T \approx 0.89$);
in both cases we consider $L = 40$, and the data set have $N_r = 10^4$ configurations.
We obtain $I_d \approx 47$ and $I_d \approx 120$, respectively. }
\label{fig2SM}
\end{figure}

Fig. \ref{fig1SM}  shows the plot of $S$ for the basic $3$-site XY example presented in the Fig. \ref{fig1} (B).
It worth mentioning that while we depict the configurations $\vec{\theta} = (\theta_1,\theta_2,\theta_3)$ for clarity of illustration in Fig. \ref{fig1} (B),
in our calculations, $\vec{\theta}$ is defined as in Eq. \eqref{confXY}.
In this way, the distance between two configurations $\vec{\theta^i}$ and $\vec{\theta^j}$: $r(\vec{\theta^i},\vec{\theta^j}) =
\sqrt{_{ 2 \sum_{k=1}^{N_s} 
\left(1 - \vec{S}_{k}^{i} \cdot \vec{S}_{k}^{j} \right)}}$,
properly takes into account the periodicity of the variables $\theta_k^{i}$.
Another important technical aspect is that the fit of $S$ is unstable for larger values of $\mu$.
As is considered in ref. \cite{Laio2017},  we discard the $10\%$ of  points characterized  by the highest values of $\mu$.
Based on this approach, we obtain $I_d \approx 1$ and $I_d \approx 3$, for  the zero and high temperature regimes, respectively, which is consistent with the value expected from physical reasonable assumptions (see Figs. \ref{fig1SM} (a1) and (a2)).

We also show some examples of the plot of  $S$  for data sets generated in the vicinity of the critical points of 
the  Ising and 2D XY models, see Fig. \ref{fig2SM} (a) and (b), respectively.
In both cases, the points $S$ are well fitted by a straight line passing through the origin.
We obtain similar results  for the other system-sizes and values of $T$ considered in this work.

In contrast, simple linear dimension reduction methods, such as Principal Component Analysis (PCA), fails to describe the $I_d $ of the XY data sets.
To illustrate this point, we employ linear PCA in the same collection of configurations considered in the last paragraph.
As can be seen from Figs. \ref{fig1SM} (b1) and (b2), even for this simple example, PCA fails to obtain the true $I_d$; for $T = 0$, $I_d^{PCA} = 2$, while for $T = 100$, $I_d^{PCA} = 6$.
This failure  is related to the fact that the XY manifolds are curved~\cite{Scheurer2019}.

However, PCA can describe the main features of the Ising data set.
Based on this, we consider the PCA estimation of the $I_d$ in Fig. \ref{fig5} of the main text.
Here, we give more details about the computation of $I_{d,PCA}$.
First, we consider the eigenvalues of the covariance matrix $\bf X^T X {\bf w}_n = \lambda_n {\bf w}_n$ (we use the same notation of ref. \cite{Scalettar2017}). 
We then define the normalized eigenvalues, $\tilde{\lambda_n} = \frac{\lambda_n}{\sum_{i}^{N_r} \lambda_i}$.
The $I_d$ is defined by choosing an \textit{ad hoc} cut-off parameter for the integrated spectrum of the covariance matrix, i.e., 
\begin{equation}
 \sum_{n=1}^{I_{d,PCA}} \tilde{\lambda}_n \approx f,
\end{equation}
where $f$ represents a fraction of the eigenvalues of the covariance matrix. 
In Fig. \ref{fig5} (c), we consider $f = 0.6$.
The value of $I_{d,PCA}$ depends on $f$. 
However, we observe that the qualitative behavior of the function $I_{d,PCA}(T)$ is not affected by the value of $f$ (as long as $f \ge 0.5$).
In particular, for all the values of $f$ that we considered (i.e., $f = 0.5,0.6,0.7,0.8$ and $0.9$), the global $I_{d,PCA}(T)$ goes to $1$ immediately below $T_c$,
and that the local $I_{d,PCA}(T)$ exhibit a non-monotonic behavior.

{\section{Scaling of the $I_d$ with the number of configurations}
\label{scalingNr}}
In this section we discuss the scaling of the $I_d$ with the number of configurations, $N_r$, considered in the data set; for all the results shown in the main text, $N_r = 5 \times 10^4$.
The first important aspect to consider is that  the TWO-NN is a  scale-dependent method.
In other words, the estimation of the $I_d$ is performed on a length scale that is related to the first and second neighbor distances of each point.
Thus,  by varying  $N_r$, one is probing a different neighborhood size like, i.e., estimating $I_d$ in different scales \cite{Laio2017}.

\begin{figure}[t]
{\centering\resizebox*{8.8cm}{!}{\includegraphics*{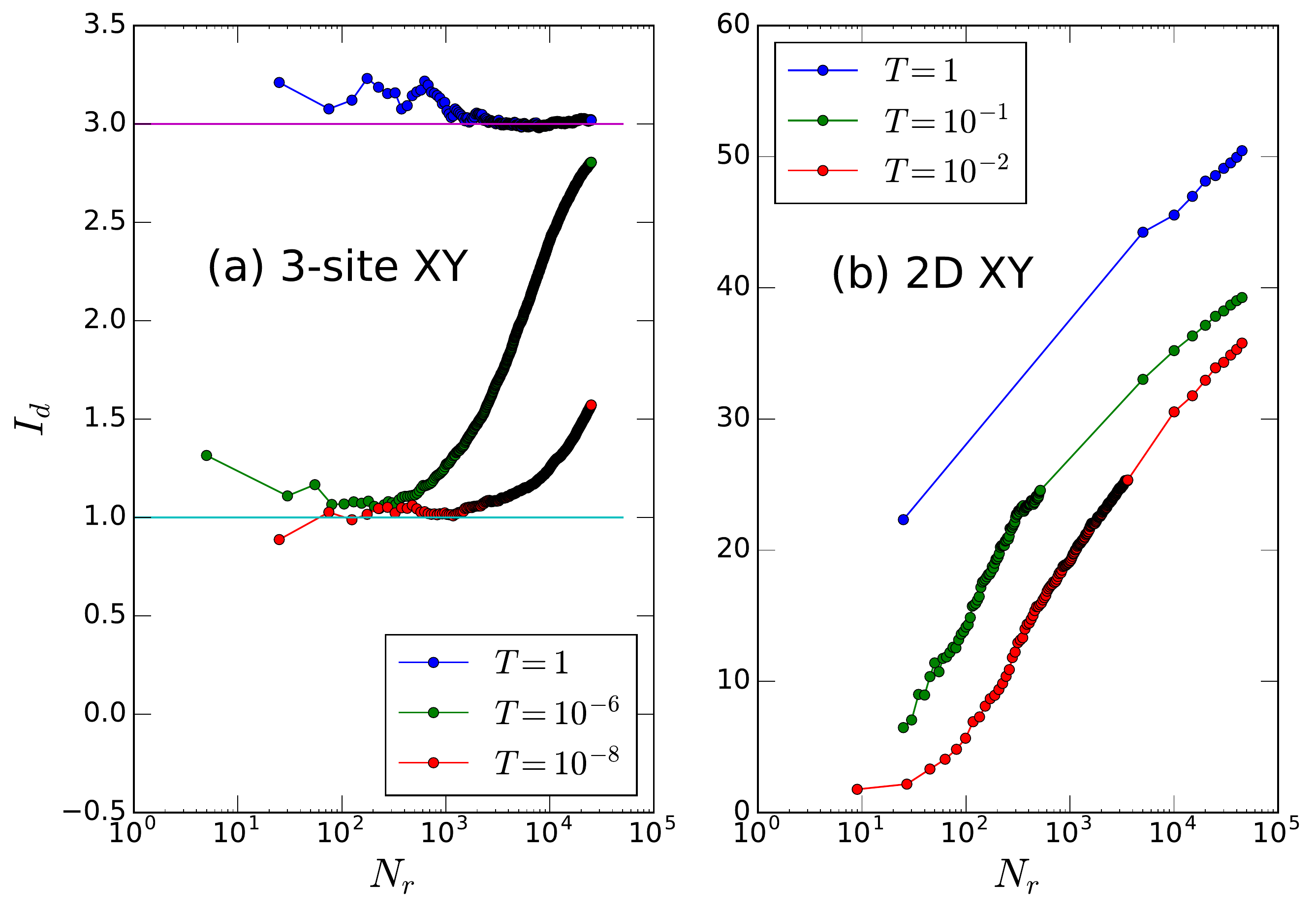}}}
\quad
\caption{Scaling of the $I_d$ with the number of configurations in the data set. In panel (a) is show the results for the $3$-site XY model,
while in panel (b) for the $2D$ XY model with $L = 10$. }
\label{fig4SM}
\end{figure}

To illustrate how this change in scale affects the  $I_d$ of the thermal data sets considered here, we first consider the $3$-site XY model in Fig. \ref{fig4SM} (a).
For $T = 1$, the $I_d$ converges to $3$ as expected for the high-temperature regime of this model.
In the low temperature regime $T \approx 10^{-6}$, however, $I_d(N_r)$ exhibit a plateau at $I_d = 1$ for $N_r \in [10^0,10^3[$. 
As illustrate in Fig.\ref{fig1} (B) this simple data set is well described by a one-dimensional manifold.
This plateau in $I_d(N_r)$ is a signature of this soft direction  \cite{Laio2017}.
Nevertheless, by further increasing $N_r$, the $I_d$ increases ($I_d \to 3$ in this case), as an effect of the decrease of the scale in which $I_d$ is estimated.
In this scale regime, the number of soft directions cannot be determined.
We stress that the computation of the $I_d$ of the  high-dimensional data sets considered here is always performed in this regime.
In this case, the $I_d$ exhibit an exponential scaling with $N_r$, as exemplified in Fig. \ref{fig4SM} (b).

\begin{figure}[t]
\centering 
\includegraphics[width=\columnwidth]{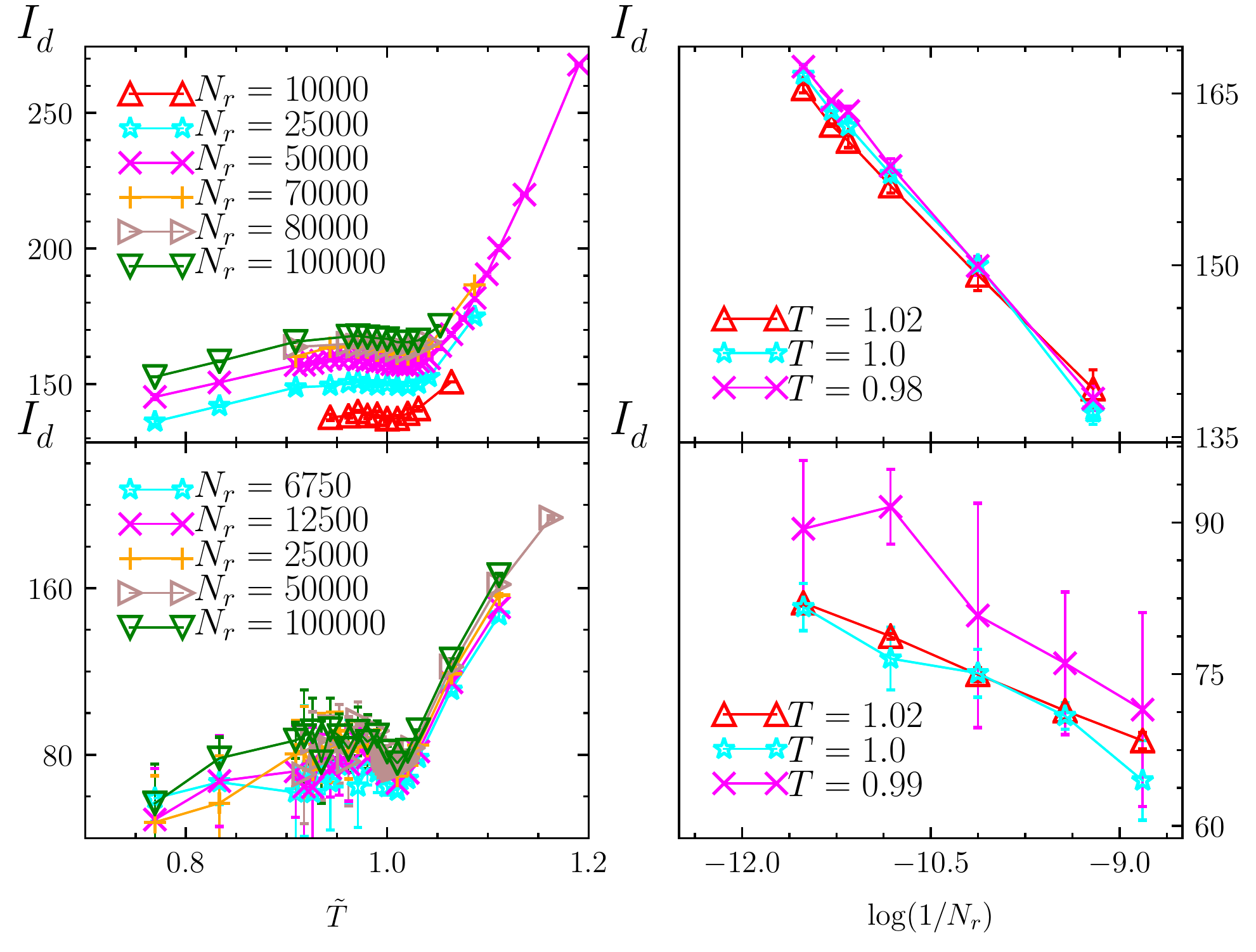}
\quad
\caption{Panels (a): the temperature dependence of the $I_d$ for different values of $N_r$ for the (a1) 2D XY model and (b) 3-states Potts model, in both cases  $L = 60$. 
For each point, we harvested approximated $10$ instances of the data set and average the resulting estimates for the $I_d$.
The error bars are the standard deviation of such set of results.
Panels (b): the scaling of $I_d$ with $N_r$ for certain values of $T$.}
\label{fig3SM}
\end{figure}

We now discuss how the temperature dependence of $I_d$ is affected by the change in $N_r$. 
Fig. \ref{fig3SM} (a1) and (b1) shows  $I_d(T)$ for different values of $N_r$ for the Potts and 2D XY data sets, respectively.
Despite the change of the absolute value of $I_d$ with $N_r$, the qualitative behavior of $I_d(T)$  is not modified.
Most importantly, we observe that the position of the local minimum at $T^*$ does not shift with $N_r$ for $N_r > 10^4$.
Furthermore, as expected, the scaling of $I_d$ with $N_r$ is exponential, see Figs. \ref{fig3SM} (a2) and (b2), at least in the vicinity of the phase transition. Similar results are obtained for other system sizes and for the Ising model. 
Summing up, our results indicate that, as long as $N_r > 10^4$, the universal scaling behavior exhibited by the $I_d$ is not affected by the scale in which the $I_d$ is measured.

\section{Data collapse}
\label{collapse}
In this section we discuss the finite size analysis employed for the estimates of the critical temperature and exponents presented in Sec.~\ref{sec:results}. 
Our procedure is a standard search of the minimal least-square difference fit between our data and an appropriately chosen scaling function hypothesis. Let us first focus on the second-order phase transitions, concerning the Ising and 3-state Potts models. The method is divided into four steps.
\begin{figure}[t]
    \centering
    \includegraphics[width=\columnwidth]{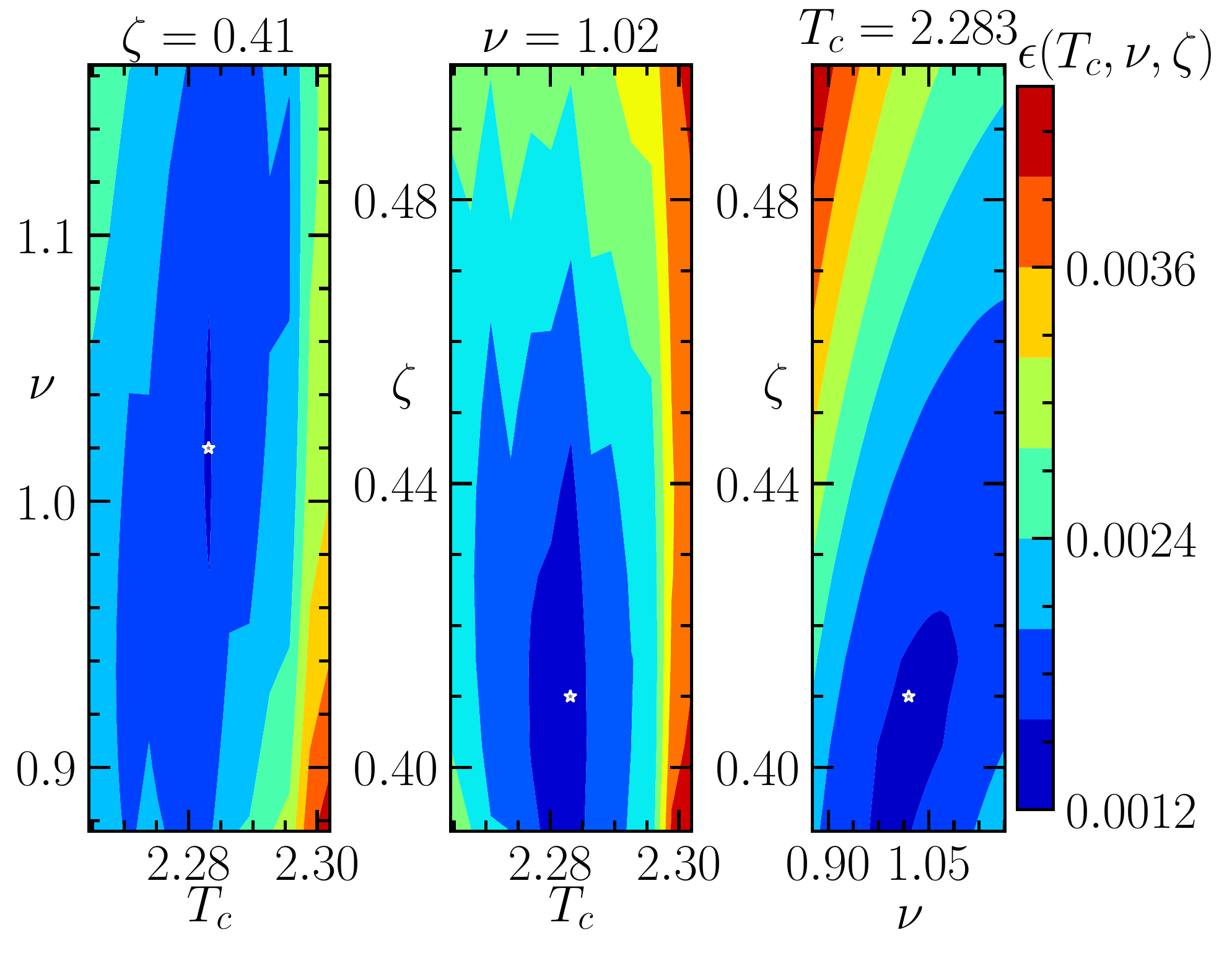}
    \caption{Contour plot of the average residuals projected on the direction of the optimal parameters for the Ising model. The white star points to the optimal parameters for our finite size analysis. }
    \label{fig:Fig8}
\end{figure}
\begin{itemize}
    \item[1] First we choose a suitable mesh for the parameter ranges for $T_c$, $\nu$ and $\alpha$.
    \item[2] We compute from our data the scaling variables $x_\textup{dat}(T_c,\nu) = (T-T_c)L^{1/\nu}$ and $y_\textup{dat}(\alpha) = I_d(T)L^{-\alpha}$ for different $R$-range of system sizes $\{L_1,L_2,\dots,L_R\}$.
    \item[3] We choose a parametric functional hypothesis $f(x;\{a\})$.
    \item[4] For each $(x_\textup{dat}(T_c,\nu),y_\textup{dat}(\alpha))$ we compute the best fit of the hypothesis function $\{a^\star\}$ through the Levenberg-Marquardt algorithm. We store the residuals as:
    \begin{equation}
        \epsilon(T_c,\nu,\alpha) = \frac{||f(x_\textup{dat}(T_c,\nu);\{a^\star\})-y_\textup{dat}(\alpha)||}{||y_\textup{dat}(\alpha)||}
    \end{equation}
\end{itemize}
The optimal set of parameters for each set $\{L_1,L_2,\dots,L_R\}$ is located in the minimum $\epsilon(T_c,\nu,\alpha)$. In order to keep a low-bias on the hypothesis function $f(x,\{a\})$, we choose various $k$-degree polynomial $Q_k(x;a_0,a_1,\dots,a_k)$. Thus, we obtain a set of optimal $\{T^\star_c\}$, $\{\nu^\star\}$, $\{\alpha^\star\}$ for each choice of degree $k$ and each set of system sizes $\{L_k\}$. Our estimates and errors for the critical temperature and critical exponents are then estimated as the average and standard deviation of these sets, respectively.

\begin{figure}[t]
    \centering
    \includegraphics[width=\columnwidth]{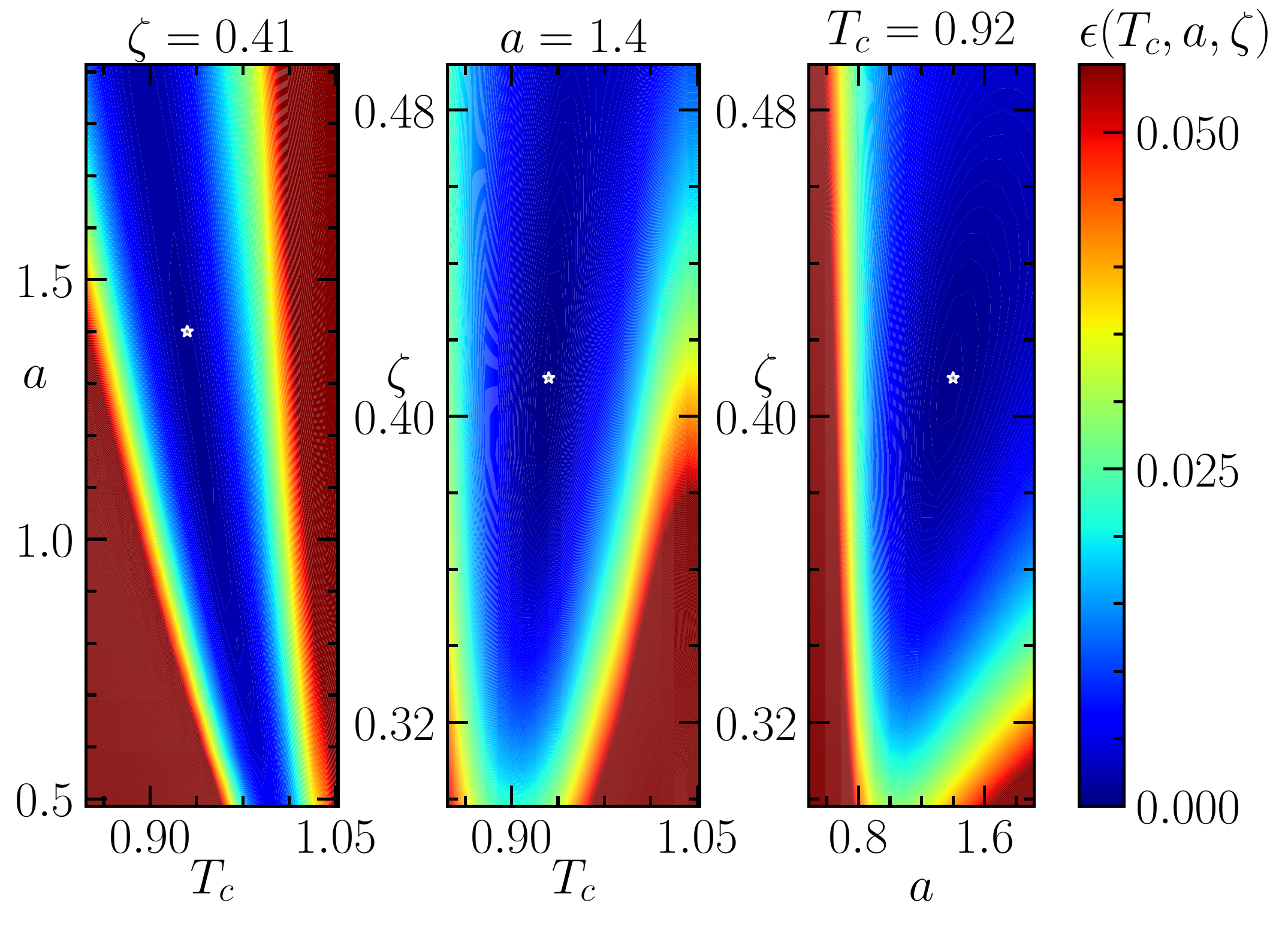}
    \caption{Contour plot of the average residuals projected on the direction of the optimal parameters for the XY model. The white star points to the optimal parameters for our finite size analysis. }
    \label{fig:Fig88}
\end{figure}

The analysis for the BKT transition (XY model) is performed in a similar fashion. The only difference is the choice of scaling variable, which for this case is:
\begin{equation}
    x_\textup{dat}(T_c,a) = L \exp\left[-\frac{a}{\sqrt{T-T_c}} \right].
\end{equation}

For the Ising, 3-state Potts and XY models we select polynomials of degrees $k\in \{5,6,7,8\}$ and different sets of system sizes among the $L\ge 70$ ones. For the XY model, we perform the data collapse within the range $T = [0.91, 1.10]$ and use a bin of $\Delta T \approx 0.005$. Our estimations for the XY model are $T_c = 0.92(1)$, $a = 1.4(1)$ and $\zeta = 0.40(1)$.

We visualize the resulting residuals for both the ising and the XY models, see Fig.~\ref{fig:Fig8} and Fig.~\ref{fig:Fig88}, respectively. Since the parameter space is 3D, for convenience we plot the projected directions along with the optimal critical parameters.

\section{Definition of the winding number}
\label{winding}

We now discuss the definition of the winding number mentioned in Fig. \ref{fig6} (b) of the main text.
We consider a closed path along the $x$ and $y$ directions of the square lattice and define
\begin{equation}
 w_x = \frac{1}{2\pi} \sum_{i=1}^{L_x} \Delta \theta_{(i,y=1)},
\end{equation}
and
\begin{equation}
 w_y = \frac{1}{2\pi}  \sum_{i=1}^{L_y} \Delta \theta_{(x=1,i)},
\end{equation}
where the angle difference is $\Delta \theta_{(x=1,i)} = \theta_{(i+1,y=1)} - \theta_{(i,y=1)}$; $\Delta \theta$ is rescaled into the range $(-\pi,\pi]$.
We compute the $w = (w_x,w_y)$ for each configuration of the data set. We then define the total number of configurations whose first two nearest neighbors have the same $w$, $N_{w}$. Figure \ref{fig6} (b) shows the fraction of those points $F_{w} = N_{w}/N_r$ as function of $T$.


\phantomsection


\begin{thebibliography}{99}
\bibitem{ft1}
We note that, by verifying this assumption, we indirectly check the ones taken before as well.
\end{thebibliography}
\end{document}